\documentclass[final]{siamltex}

\usepackage{amsmath}
\usepackage{amsfonts}
\usepackage[all]{xy}
\xyoption{color}
\xyoption{ps} 
\xyoption{dvips}
\usepackage[table,dvips,dvipsnames]{xcolor}
\usepackage{graphicx}
\usepackage{subfigure}
\def\R{\ensuremath{I\!\!R}}
\def\Rnn{\ensuremath{I\!\!R_{\geq 0}}}
\def\Rp{\ensuremath{\R_{> 0}}}
\def\YL{\ensuremath{\mathcal{Y}}}

\DeclareMathOperator{\im}{im}
\DeclareMathOperator{\sign}{sign}

\newtheorem{fact}{Fact}
\newtheorem{remark}{Remark}
\newtheorem{example}{Example}

\newcommand{\brac}[1]{\ensuremath{\left( #1 \right)}}

\definecolor{dblue}{rgb}{0.0,0.0,0.68}

\definecolor{dcyan}{rgb}{0.22,0.53,0.47}

\newcommand{\mycolors}[2]{\definecolor{#1}{rgb}{#2}}
\mycolors{shade90}{.90,.90,.90}
\mycolors{shade80}{.80,.80,.80}

\usepackage{enumerate}

\begin{document}

\title{Switching in mass action networks\\
  based on linear inequalities} 

\author{Carsten Conradi$^*$ \and Dietrich Flockerzi\thanks{Max-Planck-Institute Magdeburg, Sandtorstr.\ 1,
    Magdeburg, Germany,
    {\tt \{conradi, flockerzi\}@mpi-magdeburg.mpg.de}
  }
}

\maketitle
          


\begin{abstract}
  Many biochemical processes can successfully be described by
  dynamical systems allowing some form of switching when, depending
  on their initial conditions, solutions of the dynamical system end
  up in different regions of state space (associated with different
  biochemical functions). Switching is often realized by a bistable
  system (i.e. a dynamical system allowing two stable steady state
  solutions) and, in the majority of cases, bistability is
  established numerically. In our point of view this approach is too
  restrictive, as, one the one hand, due to predominant parameter 
  uncertainty numerical methods are generally difficult to apply to
  realistic models originating in Systems Biology. 
  And on the other hand switching already
  arises with the occurrence of a saddle type steady state
  (characterized by a Jacobian where exactly one eigenvalue is 
  positive and the remaining  eigenvalues have negative real
  part). Consequently we derive conditions based on linear
  inequalities that allow the analytic computation of states and
  parameters where the Jacobian derived from a mass action network
  has a defective zero eigenvalue so that -- under certain genericity
  conditions -- a saddle-node bifurcation occurs. Our conditions are
  applicable to general mass action networks involving at least one
  conservation relation, however, they are only sufficient (as
  infeasibility of linear  inequalities does not exclude defective
  zero eigenvalues).
\end{abstract}

\section{Introduction}
\label{sec:introduction}
Many biochemical processes can successfully be described by dynamical
systems allowing some form of switching, where, depending on, for
example, initial conditions, solutions of the dynamical system
end up in different regions of state space (associated with different
biochemical functions). Often dynamical systems admitting
bistability (i.e.\ the existence of two stable steady states) are used
for this purpose. There is a long tradition of establishing
bistability, both experimentally and computationally, in areas ranging
from signal transduction (see e.g.\ \cite{sig-023}) to cell cycle (see
e.g.\ \cite{cyc-005}).

From our point of view, however, bistability is too strong a
requirement, as already a saddle type steady state with just one
algebraically simple positive eigenvalue and all other eigenvalues
having negative real part gives rise to the desired switching
behaviour (with the global stable manifold of the saddle as a
switching surface, see \cite[Remark~3.2]{cc-flo-003}). The approach
presented here tries to directly establish such points and is hence
capable of establishing switching that is not necessarily associated
to bistability. Therefore we expect this approach to be of particular
interest for researchers working in Systems Biology and other areas of
Quantitative Biology. 

In many applications, bistability of a dynamical
system has been established numerically using bifurcation analysis or
simulations that can become arduous tasks even for relatively small
systems. Moreover parameter uncertainty is a predominant issue in
Systems Biology: the dynamical systems consist of a large number of
states and parameters, while measurement data are often very noisy and
data points and repetitions are usually few. Hence techniques allowing
the direct analytic computation of parameter vectors where a given
system exhibits switching are desirable.

In developing these techniques we identified two promising approaches:
(i) establishing multiple steady states as a mechanism for possible
switching and bistability
\cite{cc-flo-multi-002,fein-024,fein-025,cc-flo-003} and (ii)
establishing points where the dynamical system undergoes a saddle-node
bifurcation so that the global stable manifold of the saddle is acting
as a switching surface. The first approach is motivated by the so-called
Chemical Reaction Network Theory developed by Feinberg and co-workers
(see \cite{fein-013,fein-016,fein-017} and
\cite{fein-025,cc-flo-003,alg-006}). The second approach is based on
the structure of the Jacobian of a mass-action network
\cite{fein-032}. This approach was successful for a
double-phosphorylation mechanism where the nullspace of the Jacobian
admits a very special representation (cf.\, \cite{fein-032}).

Here we extend these ideas to mass action networks in general
(involving at least one conservation relation) by making use of a
property that is frequently observed in dynamical systems
originating in Systems Biology: one often faces dynamical systems
that involve so-called \emph{conservation relations} confining
trajectories to affine linear subspaces of state space.
 
Thus, the Jacobian of such a system evaluated at an arbitrary
point in state space has at least as many zero eigenvalues as there
are conservation relations. Consequently, for a saddle-node to occur
at a particular point in state space, the Jacobian has to have an
\textit{additional} zero eigenvalue at that point. Generically,
mass-action systems undergo a  bifurcation at that point -- one can
state conditions guaranteeing a saddle-node bifurcation
(cf. Section~\ref{sec:saddle-node-mass-action} or
\cite{fein-032}). One can expect that such sufficient conditions for a
saddle-node bifurcation can be established for mass action networks
originating in Systems Biology since there are many parameters which
can be chosen as continuation  parameters. Hence, such an additional
zero eigenvalue frequently entails a saddle-node bifurcation and thus
switching in a mass action network.

The main result of our paper are sufficient conditions
 guaranteeing such an additional zero eigenvalue that take the form
of linear inequality systems and are thus easy to check. Moreover, our
result is constructive in the sense that the 
solutions to one of the inequality systems determine a state and
parameter vector  where the Jacobian has an additional zero
eigenvalue and thus fulfills the  necessary degeneracy condition for a
saddle-node. Infeasibility of all inequality systems does not exclude
additional zero eigenvalues, hence feasibility of at least one
inequality system is a sufficient condition for an additional
eigenvalue. In case the remaining eigenvalues of the linearization
have negative real parts such feasibility is generically sufficient for
a saddle-node  and the associated bifurcation into a saddle and a
node. We verify this splitting in our case studies by computing
bifurcation diagrams.

Finally we'd like to point out that our results are in a certain sense
complementary to those obtained in \cite{fein-008},
\cite{fein-009,fein-019,fein-020}, \cite{fein-021},
\cite{inj-003,inj-002,inj-001}: all these references present sufficient 
conditions for the global injectivity of a dynamical system defined by
a biochemical reaction network (not necessarily restricted to mass
action systems). In particular, these conditions exclude
switching. More along the line of our work is the approach of Mincheva
and coworkers \cite{fein-044,fein-028}. There  the tight connection
between the characteristic polynomial of the Jacobian and the cycles
of certain graphs associated to the Jacobian are exploited to derive
conditions for certain instabilities (e.g.\ saddle-node or Hopf
bifurcations). The major difference to our work is that we are not
working with the characteristic polynomial but rather exploit the
fact (reported in \cite{fein-032}) that Jordan blocks of size
$\geq 2$ imply additional zero eigenvalues (and thus candidates for,
for example, saddle-node bifurcations).

In the following we briefly describe the organization of the paper and
at the same time offer the conclusions that can be drawn. In
Section~\ref{sec:dyn-sys-mass-action} we describe dynamical systems 
defined by mass action networks, recall some results from
\cite{fein-032} and \textit{characterize} positive state vectors where
the Jacobian has such an additional and thus defective zero eigenvalue
(Lemma~\ref{lem:zero_eigen_defective}, Theorem~\ref{theo:nasc}).
Those state vectors arise from elements of a semialgebraic set
that contains only polynomials of degree two or less -- regardless
of the exponents in the polynomial ODE system defined by a mass
action network. In Section~\ref{sec:suff-lin-ineq}, based on a
result from Qualitative Matrix Theory ensuring the existence of
positive null vectors, we present a sufficient condition allowing
the computation of elements of that semialgebraic set that takes the 
form of linear inequality systems. The solvability of these inequality
systems is then sufficient for the existence of an additional zero
eigenvalue (Theorem~\ref{lem:main}). In Section~\ref{sec:application},
finally, we demonstrate the applicability of the results presented
here by analyzing as a proof of principle two competing mass action
networks describing the G1/S transition in the cell cycle of budding
yeast. These networks were originally presented in \cite{fein-025} and
\cite{cc-flo-003} where their investigation was based on subnetwork
analysis. Both networks are not accessible by the results of
\cite{fein-032}.

For the convenience of the reader we provide some additional
information in four appendices. In
Appendix~\ref{sec:saddle-node-mass-action} we recall some remarks 
concerning saddle-node bifurcations in mass-action networks that
were made earlier in \cite{fein-032},  in Appendix~\ref{sec:data-tern} 
and  Appendix~\ref{sec:data-bin} we collect the relevant
structural information of the G1/S transition networks discussed in
Section~\ref{sec:application}.
The final Appendix \ref{sec:imequal}, using basic linear algebra, 
discusses  some of the assumptions and results in the present work.

\section{Dynamical systems defined by mass action systems}
\label{sec:dyn-sys-mass-action}
\mbox{}

  To introduce the notation, we use the network depicted in
  equation~(\ref{eq:exa_net}) below. This network is analysed in
  \cite{fein-013}, where multiple steady states are established.

  \begin{equation}
    \label{eq:exa_net}
    \begin{split}
      &\xymatrix{
        A+2\, S \ar @<.4ex> @{-^>}[r] ^{\bf k_1} & A\, S_2 \ar @{-^>}[l]^{\bf k_2} 
      } \\
      &\xymatrix{
        B+S \ar @<.4ex> @{-^>}[r] ^{\bf k_3} & B\, S \ar @{-^>}[l]^{\bf k_4} 
      } \\
      &\xymatrix{
        A\, S_2 + B\, S \ar [r] ^{\bf k_5} & C+3\, S 
      } \\
      &\xymatrix{
        A \ar @<.4ex> @{-^>}[r] ^{\bf k_6} & 0 \ar @{-^>}[l]^{\bf k_7} 
        \ar @<.4ex> @{-^>}[r] ^{\bf k_8} & B \ar @{-^>}[l]^{\bf k_9}  \\
        & C \ar [u] ^{\bf k_{10}}
      }
    \end{split}
  \end{equation}

  Network (\ref{eq:exa_net}) consists of $n$ \emph{species} ($n=6$)
  and with each species we associate a variable $x_i$ representing its
  concentration and the corresponding unit vector $e_i$ of $\R^6$: $x_1$ and
  $e_1$ with $A$, $x_2$  and $e_2$ with $B$,  $x_3$ and $e_3$ with
  $S$,  $x_4$ and $e_4$ with $A\, S_2$, $x_5$ and $e_5$ with $B\, S$
  and $x_6$ and $e_6$ with $C$.

  The nodes of the network graph are called \emph{complexes} and with
  each complex we associate the sum of its constituent species. The
  above network  contains $m$
  complexes ($m=10$): The complex $0$ will be denoted by the zero vector
  $0\in \R^6$ and is used to encode that the system is open with respect
  to $A$, $B$ and $C$: $A$ and $B$ can enter and leave the system
  while $C$ can only leave the system. As a complex $A$
  is associated with $e_1\in \R^6$, $B$ with $e_2$, $C$ with
  $e_6$, $A+2\, S$ with $e_1+2\, e_3$, $A\, S_2$ with $e_4$, $B+S$
  with $e_2+e_3$, $B\, S$ with $e_5$, $A\, S_2 + B\, S$ with
  $e_4+e_5$ and  $C + 3\, S$ with $e_6 + 3\, e_3$.

  The network consists of $r$ reactions ($r=10$), e.g. $A+2\, S \to
  A\, S_2$,
  where the complex at the tail of the arrow is called \emph{educt
    complex} and the complex at the tip of the arrow is called
  \emph{product complex}. To each reaction is associated a \emph{reaction
    rate} $v_i(k,x)$. For mass action systems $v_i(k,x)$ is proportional
  to the product of (powers of) concentrations of the species forming the
  educt complexes: let $y_i$ be an educt complex vector, then one has
  $v_i(k,x) = k_i\, x^{y_i}$ (where 
  $x^{p} = \prod_{j}x_j^{p_j}$ for $n$-vectors $x$ and $p$). For
  the above network one obtains
  \begin{displaymath}
    v(k,x) = 
    (\, k_{1} x_{1} x_{3}^2,\,
    k_{2} x_{4},\,
    k_{3} x_{2} x_{3},\,
    k_{4} x_{5},\,
    k_{5} x_{4} x_{5},\,
    k_{6} x_{1},\,
    k_{7},\,
    k_{8},\,
    k_{9} x_{2},\,
    k_{10} x_{6}\, )^T\ .
  \end{displaymath}
  We collect the exponents $y_i$ of the monomials contained in
  $v_i(k,x)$ in the \emph{rate-exponent matrix} $\YL$. For the above
  network one obtains the $(n\times r)$-matrix
  \begin{align*}
    \YL &=\left[y_1,\, ...\, ,y_{10}\right] \\
    &= \bigl[
    e_{1} + 2\, e_3,\,  
    e_{4},\,  
    e_{2} + e_{3},\,  
    e_{5},\,  
    e_{4} + e_5 ,\,  
    e_{1},\,  
    0,\,  
    0,\,  
    e_{2},\, 
    e_{6}
    \bigr]\ .
\end{align*}
The reactions are encoded  in the \emph{stoichiometric matrix} $S$,
  where each column corresponds to one reaction and is defined as the
  difference between product and educt complex. For example for the
  reaction $ A+2\, S \to A\, S_2 $ one obtains $r_1 = -(e_1 + 2\,e_3) +
  e_4$. The stoichiometric matrix for the above network is
  \begin{displaymath}
    S = \left[
      \begin{array}{rrrrrrrrrr}
        -1 & 1 & 0 & 0 & 0 & -1 & \phantom{-}1 & 0 & 0 & 0 \\
        0 & 0 & -1 & 1 & 0 & 0 & 0 &  \phantom{-}1 & -1 & 0 \\
        -2 & 2 & -1 & 1 & 3 & 0 & 0 & 0 & 0 & 0 \\
        1 & -1 & 0 & 0 & -1 & 0 & 0 & 0 & 0 & 0 \\
        0 & 0 & 1 & -1 & -1 & 0 & 0 & 0 & 0 & 0 \\
        0 & 0 & 0 & 0 & 1 & 0 & 0 & 0 & 0 & -1
      \end{array}
    \right]\, .
  \end{displaymath}

A reaction network then defines a dynamical system
\begin{equation}
  \label{eq:system_xdot}
  \dot x = S\, v(k,x),
\end{equation}
which in case of the above network translates to
\begin{align*}
  \dot x_1 &=  \phantom{-}k_{7}-k_{6} x_{1}-k_{1} x_{1} x_{3}^2+k_{2} x_{4} \\
  \dot x_2 &=  \phantom{-}k_{8}-k_{9} x_{2}-k_{3} x_{2} x_{3}+k_{4} x_{5} \\
  \dot x_3 &=  -k_{3} x_{2} x_{3}-2 k_{1} x_{1} x_{3}^2+2 k_{2}
  x_{4}+k_{4} x_{5}+3 k_{5} x_{4} x_{5} \\ 
  \dot x_4 &=  \phantom{-}k_{1} x_{1} x_{3}^2-k_{2} x_{4}-k_{5} x_{4} x_{5} \\
  \dot x_5 &=  \phantom{-}k_{3} x_{2} x_{3}-k_{4} x_{5}-k_{5} x_{4} x_{5} \\
  \dot x_6 &=  \phantom{-}k_{5} x_{4} x_{5}-k_{10} x_{6}\, .
\end{align*}
In general we consider a mass action network with $n$ species, $m$
complexes and $r$ reactions. Any such system defines a dynamical
system in the form given in \eqref{eq:system_xdot}. Note that $v(k,x)\in \R^r$ is
a \textbf{monomial}, \textbf{vector-valued} function of the form
\begin{displaymath}
  v(k,x) = \diag(k)\, \phi(x),
\end{displaymath}
where $\diag(k)$ is a $(r\times r)$ diagonal matrix with the $k_i$ on the 
diagonal and $\phi(x) = \left(x^{y_i}\right)_{i=1,\ldots, r}\in
\R^r$ is a vector of monomials in $x$. Note that the rate-exponent
matrix $\YL$ defined above contains the exponent vectors of the
monomials contained in $\phi(x)$. In the sequel, we speak of steady
states $(k,x)$ of (\ref{eq:system_xdot}) when $S\, v(k,x)$ vanishes
for positive $(k,x)$.

For many realistic systems in Systems Biology the matrix $S\in
\R^{n\times r}$ does not have full row rank $s := \rank(S)$ (i.e. $s <
n$). This gives rise to $n-s$ conservation relations: let $Z$ be any matrix
whose columns form a basis of $\ker(S^T)$,  the left kernel of $S$.
Solutions $x(t)$ to (\ref{eq:system_xdot}) then satisfy 
\begin{subequations}\begin{equation}
    \label{eq:system_con_rel}
    Z^T\, x(t) = Z^T\, x(0) =:c\, ,
    \end{equation}
  that is, these solutions  lie in invariant domains $x(0)+\im(S)$
  that are parallel translates of $\im(S)$. For the above example
  (\ref{eq:exa_net}) one obtains 
  \begin{equation}\label{eq:system_con_rel2}
    x_3+2 x_4+x_5 = c\ .
  \end{equation}
\end{subequations}

\subsection{The Jacobian associated to a mass action network}
\label{sec:jacobian}

At positive $(k,x)$ the Jacobian of a mass action network (by this we
mean the Jacobian of a dynamical system defined by a mass action
network) is given by:
\begin{equation}
  \label{eq:def_Jac}
  Jac(k,x) = S\, \diag\left(v(k,x)\right)\, \YL^T\,
  \diag\left(x^{-1}\right),
\end{equation}
with stoichiometric matrix $S$ and rate-exponent matrix $\YL$.

Observe that a positive pair $(k,x)$ is a steady state of
(\ref{eq:system_xdot}) if and only if $v(k,x) \in
\text{int}\left(\ker\left(S\right) \cap \Rnn^r\right)$ (where
$\text{int}(\cdot)$ denotes the relative interior). The pointed
polyhedral cone $\ker\left(S\right)\cap\Rnn^r$ is generated by a
finite set of unique (up to scalar multiplication) extreme rays
\cite{lin-004}. 
The calculation of these rays is in general
  computationally hard, however, there exists a variety of algorithms
  and software tools, for example \cite{lin-050,lin-051}.
Let $p$ be the
number of extreme rays and let $E$ be 
a matrix whose columns are generators of
$\ker\left(S\right)\cap\Rnn^r$. Then $(k,x)$ is a positive steady
state if and only if there exists a $\nu$ with
\begin{subequations}
  \begin{equation}\label{eq:def_LAM-1}
    v(k,x) = E\, \nu \, >0\, ,\quad  \nu \in \Rnn^p \, .
  \end{equation}
  So we ask for all components of $E\, \nu$ to be (strictly) positive.
  We collect all such $\nu$  in the set
  \begin{equation}
    \label{eq:def_LAM}
    {\cal V} := \left\{\nu\in\Rnn^p | E\, \nu >0 \right\}.
  \end{equation}
  Since $E$ is a nonnegative matrix, ${\cal V}$ consists of
  the positive orthant $\Rp^p$, i.e.\ the interior of $\Rnn^p $, and 
  potentially certain faces of $\Rnn^p$ (i.e.\ elements $\nu \in {\cal
    V}$ are either positive or nonnegative with predefined sign
  pattern).
  
  As we are interested in the Jacobian $Jac(k,x)$ evaluated at a
  positive steady state we use (\ref{eq:def_LAM-1}) in
  (\ref{eq:def_Jac}) to obtain
  \begin{equation}
    \label{eq:def_Jac_ss}
    Jac(k,x)\ \equiv  \ J(\nu,x) \ =  \ N(\nu)\,
    \diag\left(x^{-1}\right),\quad (\nu,x) \in
    {\cal V}\times\Rp^n\, , 
  \end{equation}
  with the $\nu$-linear
  \begin{equation}
    \label{eq:def_Jac_ssh}
    N(\nu)\ := \ S\, \diag\left(E\, \nu\right)\, \YL^T\ \ 
    \in \, \R^{n\times n}\, , \quad (\nu,x) \in
    {\cal V}\times\Rp^n\, .
  \end{equation}
\end{subequations}
We'd like to emphasize that points $(\nu,x) \in {\cal V}
\times \Rp^n$ define points $(k,x)\in\Rp^r\times \Rp^n$ via 
\begin{equation}
  \label{eq:def_k_lambda}
  k = \diag\left(\phi\left(x^{-1}\right)\right)\, E\, \nu\ .
\end{equation}
Hence finding points $(\nu,x)$ where $J(\nu,x)$ is singular
is equivalent to finding points $(k,x)$ where the Jacobian
$Jac(k,x)$ is singular.
\begin{subequations}
  Null vectors of $Jac(k,x)=J(\nu,x)$ of the form
  $\diag\left(x\right)z$ will be obtained from the identity
  \begin{equation}
    \label{eq:def_Jac_sshh}
    J(\nu,x)\diag(x)\, z\, = \, N(\nu)\, z \, = \, 
    H(z)\, \nu = 0\, ,\ \nu \in  {\cal V}\, ,
  \end{equation}
  with the $z$-linear 
  \begin{equation}
    \label{eq:def_Jac_sshg}
    H(z)\ :=\    S\diag\left(\YL^T\, z\right)E \ \in \, \R^{n\times p}\, .
  \end{equation}
  Our goal in (\ref{eq:def_Jac_sshh}) is to  
  use a  condition from Qualitative Matrix Theory that entails
  the existence of a positive null vector $\nu$ for the matrix
  $H(z)$ (cf. Theorem~\ref{lem:main}).
\end{subequations}

\begin{example}[$J(\nu,x)$ derived from network~(\ref{eq:exa_net})]
  The generator matrix of $\ker\brac{S}\cap\Rnn^r$ is given by
  \begin{subequations}
    \begin{equation}\label{eq:exa_net0}
      E = 
      \left[
        \begin{array}{ccccc}
          1&0&0&0&1 \\
          1&0&0&0&0 \\
          0&1&0&0&1 \\
          0&1&0&0&0 \\
          0&0&0&0&1 \\
          0&0&1&0&0 \\
          0&0&1&0&1 \\
          0&0&0&1&1 \\
          0&0&0&1&0 \\
          0&0&0&0&1
        \end{array}
      \right]
    \end{equation}
    and hence satisfies
    \begin{equation}\label{eq:exa_net1}
      E\, \nu >0 \Leftrightarrow \nu>0\quad \text{and thus ${\cal V}
        \equiv \Rp^5$.}
    \end{equation}
    The matrix $J(\nu,x)$ is given by
    \begin{displaymath}
      J(\nu,x) = 
      \left[
        \begin{array}{cccccc}
          -\frac{\nu_{1}+\nu_{3}+\nu_{5}}{x_{1}} & 0 & -\frac{2
            (\nu_{1}+\nu_{5})}{x_{3}} & \frac{\nu_{1}}{x_{4}} & 0 & 0 \\
          0 & -\frac{\nu_{2}+\nu_{4}+\nu_{5}}{x_{2}} &
          -\frac{\nu_{2}+\nu_{5}}{x_{3}} & 0 & \frac{\nu_{2}}{x_{5}}
          & 0 \\
          -\frac{2 (\nu_{1}+\nu_{5})}{x_{1}} &
          -\frac{\nu_{2}+\nu_{5}}{x_{2}} & -\frac{4
            \nu_{1}+\nu_{2}+5 \nu_{5}}{x_{3}} & \frac{2 \nu_{1}+3
            \nu_{5}}{x_{4}} & \frac{\nu_{2}+3 \nu_{5}}{x_{5}} & 0 \\
          \frac{\nu_{1}+\nu_{5}}{x_{1}} & 0 & \frac{2
            (\nu_{1}+\nu_{5})}{x_{3}} &
          -\frac{\nu_{1}+\nu_{5}}{x_{4}} & -\frac{\nu_{5}}{x_{5}} & 0 \\
          0 & \frac{\nu_{2}+\nu_{5}}{x_{2}} &
          \frac{\nu_{2}+\nu_{5}}{x_{3}} & -\frac{\nu_{5}}{x_{4}} &
          -\frac{\nu_{2}+\nu_{5}}{x_{5}} & 0 \\
          0 & 0 & 0 & \frac{\nu_{5}}{x_{4}} & \frac{\nu_{5}}{x_{5}}
          & -\frac{\nu_{5}}{x_{6}}
        \end{array}
      \right]\ .
    \end{displaymath}
  \end{subequations}
\end{example}

\subsection{Zero eigenvalues of the Jacobian of a mass action system}
\label{sec:zero-eigenvalues}
We assume $s=\rank(S)<n$ so that the Jacobian always has $n-s$ zero
eigenvalues. In addition we assume 
\begin{equation}
    \label{eq:imequal}
    \im (S)\, = \, \im (J(\nu ,x))\, .
    \end{equation}
In other terms, we assume the columns of the matrix $Z$ 
from (\ref{eq:system_con_rel}) to form a basis for $\ker{(J^T(\nu ,x))}$
so that $J(\nu ,x)$ does not possess more conservation laws than $S$.
In the end, we will have to validate this condition (\ref{eq:imequal})
(cf.~Appendix \ref{sec:imequal1} and \ref{sec:imequal3}).

In looking for
bifurcations, we reduce  the system to the affine subspaces
$x(0)+\im(S)$. To this end let $U$, $W$ be orthonormal bases of
$\im(S)$, $\im(S)^\perp$, respectively and introduce
\begin{subequations}\begin{align}\label{eq:reduced-system0}
  \xi &= U^T\, x,\; \eta = W^T\, x\text{ and $x(\xi,\eta) = U\, \xi + 
    W\, \eta$} \\
  \intertext{to obtain the reduced system}
  \label{eq:reduced-system}
  \dot \xi &= U^T\, S \, v\left(k, x(\xi,\eta)\right) =:
  g\left(\xi,\eta , k \right) \\
  \dot \eta &= 0.
\end{align}
\end{subequations}
Then the upper left block of the Jacobian of this mass action network
is given by
\begin{align}
  \notag
  D_\xi \, g(\xi , \eta , k) &= U^T\, Jac(k,x(\xi , \eta ))\, U\\ 
  \intertext{and at $(\nu,x)\in{\cal V} \times \Rp^n$ by}
  \label{eq:reduced_Jac_ss}
  G(\xi , \eta ,\nu ) &= 
    U^T\, J(\nu,x(\xi , \eta ))\, U\ \in \, \R^{s\times s}\, ,
\end{align}
where we recall the relation (\ref{eq:def_k_lambda})  between $k$,
$\nu$ and  $x$. In \cite{fein-032} we  presented a method that
links zero eigenvalues of $G(\xi , \eta ,\nu )$ to zero
eigenvalues of
$J(\nu,x(\xi,\eta))$. 
Lemma~\ref{lem:zero_eigen_defective} below is
  required for Theorem~\ref{theo:nasc}, the main result of this
  section. We state it here without proof, for a proof see
  \cite{fein-032}.

\vspace{2mm}
We start with some notation and, as in \cite{lin-012}, call an
eigenvalue $\lambda$ of a matrix $A\in\R^{n\times n}$ \emph{defective}
if its \emph{algebraic multiplicity} $m_{alg}(\lambda)$ is greater
than its \emph{geometric multiplicity}  $m_{geo}(\lambda)$,  that is,
if the multiplicity of $\lambda$ as a root of the characteristic
polynomial is greater than the number of linear independent
eigenvectors corresponding to $\lambda$. 
  Hence, $\lambda_0=0$ is a defective eigenvalue of 
  $A$ if and only if $\dim(\ker(A) + \im(A))$ is less $n$. 
  This can be stated in the following way:

\vspace{2mm}\begin{fact}
  \label{fact:x_in_im_ker}
  $\lambda_0 = 0$ is a defective eigenvalue of a matrix $A\in\R^{n\times n}$
  iff there exists an $x\neq 0$ with $x\in \im\left(A\right)\cap
  \ker\left( A \right)$.
\end{fact}

  \begin{remark}
    An alternative argument for Fact~\ref{fact:x_in_im_ker}
     is based on the Jordan Canonical Form of a matrix $A$ (cf.,
    for example, \cite{lin-012}). Assume an $n \times n $ matrix $A$
    with eigenvalue $\lambda_0=0$ and $m_{alg}(\lambda_0) >
    m_{geo}(\lambda_0)$ in Jordan Canonical Form. Then the $m_{alg}
    \times m_{alg}$ block matrix corresponding to $\lambda_0$ is 
    not the zero-matrix, implying the existence of nontrivial $u_1\neq u_2$ with
    $A\, u_1 = 0$ and $A\, u_2=u_1$ and hence $u_1 \in \ker(A) \cap
    \im(A)$.
  \end{remark}

\vspace{2mm}
\noindent We recall another fact from Lemma~1 in \cite{fein-032}:

\begin{lemma} 
  \label{lem:zero_eigen_defective}
  Let $A\in\R^{n\times n}$ be a matrix of rank $s<n$ 
  and let $U$ be orthonormal basis for $\im \left(A\right)$.
  Then $\lambda_0=0$ is a defective eigenvalue of $A$ if and only if
  $\lambda_0 = 0$ is an eigenvalue of $B_1:= U^T\, A\, U \in \R^{s\times s}$.
\end{lemma}

\vspace{2mm}\noindent Based on Fact~\ref{fact:x_in_im_ker} and
Lemma~\ref{lem:zero_eigen_defective} one is led to following
observation:

\vspace{2mm}\begin{lemma}
  \label{fact:zero-eigenvalues-reduced}Let $Z_0$ be a basis
    of $\im\brac{S}^\perp$. Then 
     the Jacobian $G(\xi , \eta ,\nu )$ of the reduced system,
    evaluated at $\nu \in {\cal V}$ and $x=U\, \xi + W\, \eta \in \Rp^n$ (cf.\
    (\ref{eq:reduced_Jac_ss}) and (\ref{eq:reduced-system0})),
    has a zero eigenvalue if and only if there
  exist a nontrivial vector $z\in\R^n$, a vector $x\in\Rp^n$
  and a vector $\nu\in{\cal V}$ with
  \begin{subequations}
    \begin{align}
      \label{eq:bif_1}
      H(z)\, \nu &= 0 \\
      \label{eq:bif_2}
      Z_0^T\, \diag(x)\, z &= 0.
    \end{align}
  \end{subequations}
\end{lemma}
In the sequel, we take for $Z_0$ the matrix $Z$ describing the conservation laws
(cf. (\ref{eq:system_con_rel})).

  \begin{proof}
    From Lemma~\ref{lem:zero_eigen_defective}
    follows that $G(\xi , \eta, \nu )$ has $\lambda_0=0$ as an
    eigenvalue, if and only if $J(\nu,x)$ has $\lambda_0=0$ as a
    defective eigenvalue. From Fact~\ref{fact:x_in_im_ker} follows
    that $J(\nu,x)$ has a defective eigenvalue, if and only if
    there is a  nontrivial vector $\tilde z \in\ker\brac{S}\cap\im\brac{S}$. That
    is, $\tilde z$ must satisfy $N(\nu)\, \diag\brac{\frac{1}{x}}\,
    \tilde z = 0$ and $Z^T\, \tilde z = 0$ (cf.\ (\ref{eq:def_Jac_ss})
    and (\ref{eq:imequal})). Let $\tilde
    z=\diag\brac{x}\, z$, then \eqref{eq:bif_1} and \eqref{eq:bif_2}
    follow. 
  \end{proof}

\vspace{2mm}First we consider condition~(\ref{eq:bif_2}) and establish necessary
and sufficient conditions for the existence of solutions ($x$, $z$)
$\in \Rp^n\times\R^n$, where we assume that $z$ is given.

\vspace{2mm}\begin{lemma}
  \label{sec:lem_pos_x_sign_z}
  Let $M\in R^{q_1\times q_2}$ be any matrix and let $z\in\R^{q_1}$ be given. Then
  there exists a positive vector $x\in\Rp^{q_1}$ such that 
  \begin{displaymath}
    M^T\, \diag(x)\, z = 0,
  \end{displaymath}
  if and only if 
  \begin{equation}
    \label{eq:omega_condi}
    \exists \omega\in\ker\left(M^T\right)\; \text{with
      $\sign(\omega)=\sign(z)$.}
  \end{equation}
  In this case $x=\left(x\right)_{i=1,\, \ldots, q_1}$ is given by
  \begin{equation}
    \label{eq:x_def}
    x_i =
    \begin{cases}
      \frac{\omega_i}{z_i}\text{, if $z_i\neq 0$} \\
      \bar x_i > 0\text{, arbitrary, if $z_i=0$.}
    \end{cases}
  \end{equation}
\end{lemma}
\begin{proof}
  Assume $M^T\, \diag(x)\, z=0$ holds for positive $x$ and some
  $z$. Then $\omega:= \diag(x)\, z \in\ker\left(M^T\right)$ and
  $\sign(\omega)=\sign(z)$ follows from positivity of $x$. Vice versa,
  let $z\in\R$ and $\omega\in\ker\left(M^T\right)$ with
  $\sign(\omega)=\sign(z)$ be given. Let $x$ be as in
  (\ref{eq:x_def}). Then $\sign(\omega)=\sign(z)$ implies positivity
  of $x$ and one has $\diag(x)\, z = \omega \in \ker\left(M^T\right)$.
\end{proof}

\begin{remark}
  Observe that, given a vector $z$, the condition (\ref{eq:omega_condi})
  takes the form of linear inequalities: one has to establish
  feasibility of the system
  \begin{displaymath}
    M^T\, \omega = 0\text{, $\sign\left(z_i\right)\, \omega_i > 0$, if
    $z_i\neq 0$ and $\omega_i=0$, if $z_i = 0$.}
  \end{displaymath}
\end{remark}

\begin{remark}[Connection to \cite{fein-032}]
  The condition (\ref{eq:bif_1}) requires the \textbf{symbolic
    computation} of $\ker\left(H(\nu)\right)$. This can be of
  forbidding complexity, especially for large 
  networks, even though it is in principle possible. \\
  So far, the only application of the simple fact in
  Lemma~\ref{lem:zero_eigen_defective}
  we are aware of was in \cite{fein-032}. There we analysed a mass
  action network describing the double phosphorylation of a
  protein. For this network we obtained a symbolic representation
  of $\ker\left(N(\nu)\right)$ that could be brought into a
  $\nu$-independent form. In general, the previous approach requires
  positive solutions to some well-defined polynomial equations in
  $\nu$ and is thus limited to  certain classes of systems (cf.\
  \cite{cc-flo-multi-002}).
\end{remark}


\vspace*{2mm}
In the sequel, we employ the structure of $H(z)$, given by
(\ref{eq:def_Jac_sshg}), when discussing $H(z)\, \nu = 0$.
Let the columns of $S_0\in\R^{r\times (r-s)}$ be a basis of $\ker(S)$
and let $S_\#\in\R^{r\times r}$ be a matrix such that 
$S_\#\, S_0 = \left[
  \begin{smallmatrix}
    I_{r-s} \\ {\bf 0}_{s\times (r-s)}
  \end{smallmatrix}
\right]$.
If we let the columns of $S_c$ be a basis for $\im(S^T)$ and if we denote the Moore-Penrose
inverse $(S_0^TS_0)^{-1}S_0^T$ by $S_0^\#$ 
we will consider a particular such $S_\#$ by setting $S_\#^{part}=\left[
  \begin{smallmatrix}
     S_0^\#\\ S_c^T
  \end{smallmatrix}
\right]$.

Equation~(\ref{eq:bif_1}) is now equivalent to
\begin{equation}\label{eq:Ssharp0}
S_0\, \alpha \, =\, \diag\left(\YL^T\, z\right)\, E\, \nu 
\end{equation}
for some vector $\alpha \in \R^{r-s}$ and, by left multiplication with
$S_\#^{part}$,  to 
\begin{equation}
  \label{eq:Teq}
  \left[
    \begin{array}{c}
      \alpha\\0
    \end{array}
  \right] = \left[\begin{array}{c}
                            P(z) \\ Q(z)
                           \end{array}
                     \right]\, \nu\, \quad \mbox{with} \ 
  \left[
    \begin{array}{c}
      P(z) \\ Q(z)
    \end{array}
  \right] :=S_\#^{part} \diag\left(\YL^T\, z\right)E\, .
\end{equation}
Observe that $z$ and $\nu$ satisfy $H(z)\, \nu = 0$
(cf.(\ref{eq:def_Jac_sshh})\&(\ref{eq:def_Jac_sshg}))
if and only if one
has 
\begin{equation}
  \label{eq:Qz_condi}
  Q(z)\, \nu = 0,\quad \nu \in {\cal V}\, .
\end{equation}
The corresponding $\alpha$ will be given by $P(z)\nu$.
We note that the elements of the matrices $P(z)$ and $Q(z)$
are linear forms in $z$. Appendix~\ref{sec:imequal2} shows that the
condition (\ref{eq:Qz_condi}) is independent from the chosen bases for
$\ker(S)$ and $\im(S^T)$.

\vspace{2mm}\begin{theorem} 
  \label{theo:nasc}
  The Jacobian $G(\xi , \eta ,\nu)$ of the reduced system, 
  evaluated at $\xi$ and $\eta$ with
  $x=x(\xi,\eta)\in \Rp^n$ as in (\ref{eq:reduced-system0}) and $\nu \in {\cal V}$,
  has zero
  as an eigenvalue with algebraic multiplicity $\geq 1$ if and only if there
  exist $z\in \R^n$, $\omega\in \R^n$ and $\mu \in \Rnn^p$ with  
  \begin{equation}\label{eq:nasc}  
    Q(z)\mu =0,\ \ E\mu >0,\ \ Z^T\omega=0,\ \ \sign(\omega) = \sign(z)\, .\end{equation}  
\end{theorem}
\begin{proof}
  With the settings $\omega=\diag(x)z$ as in (\ref{eq:x_def}), 
  $x=U\xi+W\eta$  as in (\ref{eq:reduced-system0}) and $\nu=\mu$,
  Theorem~\ref{theo:nasc} follows immediately from the 
  Lemma~\ref{fact:zero-eigenvalues-reduced}, Lemma~\ref{sec:lem_pos_x_sign_z} and 
  the equivalence of $H(z)\nu=0$ with (\ref{eq:Qz_condi}).
\end{proof}

\vspace{2mm}\begin{remark}[Open condition (\ref{eq:Teq})]\label{openeq:Teq}
Observe that $Q(z)\, \mu = 0$ in (\ref{eq:nasc}) can also be written in the form 
$\tilde{Q}(\mu)\, z=0$ 
since $N(\mu)z=H(z)\mu$ (cf.(\ref{eq:def_Jac_sshh}) with (\ref{eq:def_Jac_ssh})
\& (\ref{eq:def_Jac_sshg})) implies the equivalence of
(\ref{eq:Teq}) and
\begin{equation}
  \label{eq:Teq1}
  \left[
    \begin{array}{c}
      \alpha\\0
    \end{array}
  \right] = \left[\begin{array}{c}
                            \tilde{P}(\mu) \\ \tilde{Q}(\mu)
                           \end{array}
                     \right]\, z\, \quad \mbox{with} \ 
  \left[
    \begin{array}{c}\tilde{P}(\mu) \\ \tilde{Q}(\mu)\end{array}
  \right] :=S_\#^{part} \diag\left(E\, \mu\right)\YL^T\ .
\end{equation}
This reformulation reveals that  (\ref{eq:Teq}) is
an \textit{open} condition: Given a particular solution $(\tilde{z},\tilde{\omega},\tilde{\mu})$,
there will exist a solution $(z,\omega,\mu)$ for all $\mu$'s that are sufficiently close to
$\tilde{\mu}$. So, there is some freedom in the choice of $\mu$,
cf.~Appendix \ref{sec:imequal3}.
\end{remark}

\vspace{2mm}Note that the semialgebraic set given by (\ref{eq:nasc}) is always
  defined by polynomials of degree two or less, 
  independent of the exponents in the polynomial ODEs. Any
  element gives rise to a defective eigenvalue $0$ of the Jacobian
  $Jac(k,x)$. For the computation of elements of that set we will
  later on employ the following observation:
in case the vector $\mu$ in (\ref{eq:nasc}) can be chosen as a
positive null vector of $Q(z)$, the condition $E\mu>0$ is
automatically satisfied.  Thus we arrive at a sufficient condition for
a defective eigenvalue $0$ of the Jacobian $Jac(k,x)$ by  
imposing conditions on the matrix  $Q(z)$ that imply the existence of
a positive null vector $\mu$ and conditions on the vector $z$ ensuring
the sign-compatibility of $z$ with $\ker{(Z^T)}$.

Since the elements of
$Q(z)$ derived from a mass action network are always linear forms in
$z$, one can determine all sign patterns that
$\sign\brac{Q(z)}$ can admit by analyzing the corresponding
inequality systems. The idea is to look for sign patterns
guaranteeing that \emph{every matrix} with that sign pattern has a
\emph{positive} kernel vector. To this end we resort in subsection \ref{sec:new_condi} to
\emph{Qualitative Matrix Theory} \cite{sign-009} and to
\emph{$L^+$-matrices} in particular \cite{sign-004}.
We first exemplify our approach by examining (\ref{eq:nasc}) for
network~(\ref{eq:exa_net}) and turn to the general case in
Section~\ref{sec:new_condi}.

\vspace{3mm}\section{Conditions for a singular  reduced Jacobian $G$}
\label{sec:suff-lin-ineq}\mbox{}


\subsection{System (\ref{eq:nasc}) for  network~(\ref{eq:exa_net})}
Note that for network~(\ref{eq:exa_net}) the matrix $E$ of (\ref{eq:exa_net0})
is also a
basis for $\ker\brac{S}$ (in general this need not be the
case). Using  this $E$ we obtain for equation (\ref{eq:Teq})
(where gray indicates rows belonging to $Q(z)$):
\begin{displaymath}
  \left[
    \begin{array}{ccccc} 
      z_{4} & 0 & 0 & 0 & 0 \\
      0 & z_{5} & 0 & 0 & 0 \\
      0 & 0 & 0 & 0 & -z_{6} \\
      0 & 0 & 0 & z_{2} & 0 \\
      0 & 0 & 0 & 0 & z_{6} \\ \hline
      \rowcolor{shade80} z_{1}+2 z_{3}-z_{4} & 0 & 0 & 0 & z_{1}+2 z_{3}-z_{6} \\ 
      \rowcolor{shade80} 0 & z_{2}+z_{3}-z_{5} & 0 & 0 & z_{2}+z_{3}-z_{6} \\
      \rowcolor{shade80} 0 & 0 & 0 & 0 & z_{4}+z_{5}-z_{6} \\
      \rowcolor{shade80} 0 & 0 & z_{1} & 0 & z_{6} \\
      \rowcolor{shade80} 0 & 0 & 0 & -z_{2} & -z_{6}
    \end{array}
  \right]\, \nu = \left[
    \begin{array}{c}
      \alpha_1 \\ 
      \\ 
      \vdots \\
      \\ 
      \alpha_5 \\ \hline
      \rowcolor{shade80} 0 \\
      \rowcolor{shade80} \\ 
      \rowcolor{shade80} \vdots  \\
      \rowcolor{shade80}\\  
      \rowcolor{shade80} 0
    \end{array}
  \right]\ .
\end{displaymath}
One has $s=5$ and $r=10$, hence the matrix $Q(z)$ is defined by rows
6--10. However, it is easy to see that $v\in {\cal V}$ (and hence
positive $\nu$ by (\ref{eq:exa_net1})) exist only if
$z_6=z_4+z_5$. Hence $Q(z)$ consists
only of the rows 6, 7, 9 and 10 as row 8 evaluated at $z_6=z_4+z_5$ is
identically zero. One obtains
\begin{displaymath}
  Q(z) = 
  \left[
    \begin{array}{ccccc}
      z_{1}+2 z_{3}-z_{4} & 0 & 0 & 0 & z_{1}+2 z_{3}-z_{4}-z_{5} \\
      0 & z_{2}+z_{3}-z_{5} & 0 & 0 & z_{2}+z_{3}-z_{4}-z_{5} \\
      0 & 0 & z_{1} & 0 & z_{4}+z_{5} \\
      0 & 0 & 0 & -z_{2} & -z_{4}-z_{5}
    \end{array}
  \right]
\end{displaymath}
For this $Q(z)$ one has positive $\nu$, iff the following pairs of
linear forms are either of opposite sign or both equal to zero:
\begin{equation}
  \label{eq:linear_forms_exa1}
  \begin{split}
    \ell_1(z) := z_1+2 z_3-z_4\quad &\text{and}\quad  \ell_2(z) :=
    z_1+2 z_3-z_4-z_5,\\ 
    \ell_3(z) := z_2+z_3-z_5\quad &\text{and}\quad  \ell_4(z) :=
    z_2+z_3-z_4-z_5, \\
    \ell_5(z) := z_1\quad &\text{and}\quad \ell_6(z) := z_4+z_5, \\
    \ell_7(z) := -z_2\quad &\text{and}\quad \ell_8(z):=
    -z_4-z_5\ .
  \end{split}
\end{equation}
These conditions can be expressed as \emph{linear inequality systems},
for example 
\begin{subequations}
  \begin{equation}
    \label{eq:z_ineq_exa1_1}
    \begin{split}
      z_1+2 z_3-z_4>0,\,  z_1+2 z_3-z_4-z_5<0,\\
      z_2+z_3-z_5<0,\, z_2+z_3-z_4-z_5>0,\\
      z_1>0,\,  z_4+z_5<0,\\
      -z_2<0,\,  -z_4-z_5>0.
    \end{split}
  \end{equation}
  This system is feasible; pick any $\tilde z\in\R^5$ satisfying
  (\ref{eq:z_ineq_exa1_1}) and let $\tilde z_6=\tilde z_4 + \tilde
  z_5$. Then $Q(z)$ evaluated at that $\tilde z$ has a positive
  kernel vector $\nu$ (cf. Table~\ref{tab:z_w_exa1} and
  (\ref{eq:nu14}).\\ 
  We apply  Lemma~\ref{sec:lem_pos_x_sign_z} with
  $M^T=Z^T=\left(0,\, 0,\, 1,\, 2,\, 1,\, 0\right)$ from
  (\ref{eq:system_con_rel2}) and need to find a vector
  $\tilde\omega\in\ker\brac{Z^T}$ with $\sign\brac{\tilde\omega} =
  \sign\brac{\tilde{z}}$. For the choice of 
  $\tilde\omega$ with $\tilde\omega_3<0$, $\tilde\omega_4<0$ and
  $\tilde\omega_5>0$ we consequently  add
  \begin{equation} 
    \label{eq:z_ineq_exa1_2}
    \tilde{z}_3<0,\, \tilde{z}_4<0,\, \tilde{z}_5>0
  \end{equation}
\end{subequations}
to the inequality system. The overall system (\ref{eq:z_ineq_exa1_1})
\& (\ref{eq:z_ineq_exa1_2}) is feasible and one solution $\tilde z$ is
given in Table~\ref{tab:z_w_exa1}. Table~\ref{tab:z_w_exa1} also
contains a vector $\tilde\omega\in\ker\brac{Z^T}$ with
$\sign\brac{\tilde z} = \sign\brac{\tilde\omega}$ and the vector
$x=\frac{\tilde\omega}{\tilde z}$ (cf.\
Lemma~\ref{sec:lem_pos_x_sign_z}, equation (\ref{eq:x_def})).
\begin{table}[htb]
  \centering
  \begin{tabular}{|>{$}c<{$}|*{6}{>{$}r<{$}}|}\hline
    & \tiny 1 & \tiny 2 & \tiny 3 & \tiny 4 & \tiny 5 & \tiny 6 \\ \hline
    \tilde z& 4 & 1 & -5 & -8 & 3 & -5 \\
    \tilde \omega & 1 & 1 & -3 & -1 & 5 & -1 \\
    x & \frac{1}{4} & 1 & \frac{3}{5} & \frac{1}{8} & \frac{5}{3} &
    \frac{1}{5} \\ \hline
  \end{tabular}
  \caption{Vectors $\tilde{z}$, $\tilde{\omega}$ and $x$}
  \label{tab:z_w_exa1}
\end{table}
Evaluating $Q(z)$ at $\tilde z$ from Table~\ref{tab:z_w_exa1} one has the
matrix
\begin{displaymath}
  Q(\tilde{z}) = \left[
    \begin{array}{rrrrr}
      2 & 0 & 0 & 0 & -1 \\
      0 & -7 & 0 & 0 & 1 \\
      0 & 0 & \phantom{-}4 & 0 & -5 \\
      0 & 0 & 0 & -1 & 5
    \end{array}
  \right]
\end{displaymath}
that has the positive kernel vector
\begin{equation}\label{eq:nu14}
  \nu = 
  \left(
    14,\, 
    4,\, 
    35,\, 
    140,\,
    28
  \right)^T\ .
\end{equation}
Vector $x$ from Table~\ref{tab:z_w_exa1} and the above $\nu$
define a vector of rate constants:
\begin{displaymath}
  k = \left(
    \frac{1400}{3},\, 
    112,\, 
    \frac{160}{3},\, 
    \frac{12}{5},\, 
    \frac{672}{5},\, 
    140,\, 
    63,\, 
    168,\, 
    140,\,
    140
  \right)^T\ .
\end{displaymath}
Evaluation of $Jac\brac{k,x}$ at this $k$ and $x$ from
Table~\ref{tab:z_w_exa1} confirms $\lambda=0$ as a defective
eigenvalue. 

All in all there are 81 different inequality systems where the
pairs from (\ref{eq:linear_forms_exa1}) are of different sign or
both zero. There are also 13 inequality systems
like~(\ref{eq:z_ineq_exa1_2}) that constrain $z$ such that there
is a $\omega\in\ker\brac{Z^T}$ with $\sign\brac{\omega} =
\sign\brac{z}$. Of these 13*81=1053 inequality systems only the
following four are feasible:

\begin{minipage}[t]{0.48\linewidth}
  \begin{align*}
    z_3<0&,\; z_4<0,\; z_5>0 \\
    \ell_1(z) &>0,\; \ell_2(z) <0 \\
    \tag{$P_1^+$}\label{eq:P1}
    \ell_3(z) &<0,\; \ell_4(z) >0 \\ 
    \ell_5(z) &>0,\; \ell_6(z) <0 \\
    \ell_7(z) &<0,\; \ell_8(z) >0 \\
    \intertext{and}
    z_3>0&,\; z_4>0,\; z_5<0 \\
    \ell_1(z) &<0,\; \ell_2(z) >0 \\
    \tag{$P_1^-$}\label{eq:P3}
    \ell_3(z) &>0,\; \ell_4(z) <0 \\
    \ell_5(z) &<0,\; \ell_6(z) >0 \\
    \ell_7(z) &>0,\; \ell_8(z) <0 
  \end{align*}
\end{minipage}
\begin{minipage}[t]{0.48\linewidth}
  \begin{align*}
    z_3<0&,\; z_4>0,\; z_5<0 \\
    \ell_1(z)&<0,\; \ell_2(z) > 0 \\
    \tag{$P_2^+$}\label{eq:P2}
    \ell_3(z)&>0,\; \ell_4(z) < 0 \\
    \ell_5(z)&>0,\; \ell_6(z) < 0 \\
    \ell_7(z)&<0,\; \ell_8(z) > 0 \\
    \intertext{\phantom{and}}
    z_3>0&,\; z_4<0,\; z_5>0 \\
    \ell_1(z)&>0,\; \ell_2(z) < 0 \\
    \tag{$P_2^-$}\label{eq:P4}
    \ell_3(z)&<0,\; \ell_4(z) > 0 \\
    \ell_5(z)&<0,\; \ell_6(z) > 0 \\
    \ell_7(z)&>0,\; \ell_8(z) < 0
  \end{align*}
\end{minipage}

\vspace{2mm}\noindent Because of the definitions of $\ell_5,\ell_6$
and $\ell_7$ in (\ref{eq:linear_forms_exa1}), feasible $z$'s do not
have vanishing components. All in all we have established the
following necessary and sufficient condition for a defective
eigenvalue of $J(\nu,x)$ of (\ref{eq:exa_net}). 

\vspace{2mm}  \begin{fact}
  The Jacobian $J(\nu,x)$ of (\ref{eq:exa_net}) evaluated at
  $(\nu,x) \in {\cal V} \times \Rp^6$ (and hence $Jac(k,x)$
  evaluated at positive $(k,x)$ via (\ref{eq:def_k_lambda})) has
  $\lambda_0=0$ as a defective eigenvalue, if and only if $\nu$
  and $x$ satisfy:
  \begin{enumerate}
  \item The vector $x$ can be written as $x = \frac{
      \omega}{z}$ with 
    (i) $ z \in \R^6$ satisfies one of the inequality systems
    ($P_1^\pm$), ($P_2^\pm$) and $z_6=z_4 + z_5$ (implying ${z}_i\neq
    0$ for $i=1,...,6$),\, 
    (ii) $\omega\in\ker\brac{Z^T}$ and (iii) $\sign\brac{z} =
    \sign\brac{\omega}$. 
  \item The above $z$ and the vector $\nu>0$ are such that 
    $Q(z)\, \nu = 0$.
  \end{enumerate}
\end{fact}

\vspace{2mm}\noindent  Note that, if $z\in\R^6$ with $z_6=z_4+z_5$
satisfies one of the systems ($P_1^\pm$) and ($P_2^\pm$),
then the sign
pattern $\sign\brac{Q(z)}$ is one of the following:
\begin{itemize}
\item If $z$ satisfies ($P_1^\pm$) then
  \begin{displaymath}
    \sign\brac{Q(z)} = \pm \, \left[
      \begin{array}{rrrrr}
        \phantom{-} 1 &  0 &  0 &  0 & -1 \\
        0 & -1 &  0 &  0 &  1 \\
        0 &  0 & \phantom{-} 1 &  0 & -1 \\
        0 &  0 &  0 & -1 &  1
      \end{array}
    \right]\ .
  \end{displaymath}
\item If $z$ satisfies ($P_2^\pm$) then
  \begin{displaymath}
    \sign\brac{Q(z)} =\pm \,  \left[
      \begin{array}{rrrrr}
        -1 &  0 &  0 &  0 &  1 \\
        0 &  \phantom{-}1 &  0 &  0 & -1 \\
        0 &  0 &  \phantom{-}1 &  0 & -1 \\
        0 &  0 &  0 &  -1 & 1
      \end{array}
    \right]\, .
  \end{displaymath}
\end{itemize}

\subsection{A sufficient condition}
\label{sec:new_condi}

For the example of network~(\ref{eq:exa_net}) we obtained necessary and
sufficient conditions in  form of linear inequalities in  $z$
guaranteeing a \emph{positive} kernel vector $\nu $ of $Q(z)$. 
The idea is to look for sign patterns $\sign\brac{Q(z)}$
guaranteeing that \emph{every matrix} with that sign pattern has a
\emph{positive} kernel vector (as it has been the case with the sign
patterns of the previous section). 
By \emph{Qualitative Matrix Theory} \cite{sign-009} (see in particular \cite{sign-004})
one has the following Theorem`\ref{theo:L+-matrices}. In our application,
it can be stated in the following way: If a sign pattern
$\sign\brac{Q(z)}$ is an $L^+$-matrix, then every matrix with the same
sign pattern has a positive kernel vector.

\vspace{2mm}\begin{theorem}[cf. \cite{sign-004}, Theorem 2.4, p.6]
  \label{theo:L+-matrices}
  For a $(m\times n)$ sign pattern $A$, the following are equivalent:
  \begin{enumerate}[{(}a{)}]
  \item $A$ is an $L^+$-matrix.
  \item Every matrix with the sign pattern $A$ has a \underline{positive null
    vector} and $A$ has no zero row.
  \item For each nonzero vector $\sigma\in\left\{-1,\, 0,\,
  1\right\}^{m}$, some column of \,
    $\diag\left(\sigma\right)A$\, is nonzero and nonnegative.
  \item For each nonzero vector $\sigma\in\left\{-1,\, 0,\,
  1\right\}^{m}$, some column of \,
    $\diag\left(\sigma\right)A$\, is nonzero and nonpositive.
  \end{enumerate}
\end{theorem}
Note that Theorem~\ref{theo:L+-matrices} already contains -- by the
parts $(c)$ or $(d)$ -- a primitive algorithm to determine whether or
not a given sign pattern is an $L^+$-matrix. So by
Theorem~\ref{theo:L+-matrices} one can decide whether or not a
particular sign pattern is an $L^+$-matrix.

\vspace{2mm} With respect to the $z$-linear matrix 
$Q(z)=Q_{ij}(z)\in \R^{s\times p}$ from
(\ref{eq:Qz_condi}) we propose the following:  We first stack the columns of $Q$
and consider the column vector 
\begin{displaymath}
  (Q_{11},\ldots,Q_{s1},\ldots\ldots ,Q_{1p},\ldots,Q_{sp})^T\, .
\end{displaymath}
Then we omit the components that are trivial linear forms to obtain a
bijective mapping  of the form 
\begin{equation}\label{map-vec}
\psi:\ Q(z)\in \R^{s\times p} \ \mapsto \ L\, z=(\ell_1z,\ldots \ldots ,\ell_\gamma z)^T\in \R^\gamma
\end{equation}
with nontrivial $n$-dimensional row-vectors $\ell_1$, \ldots,
$\ell_\gamma$, $\gamma \leq sp$. So, the $(\gamma
\times n)$-matrix 
\begin{equation}
  \label{eq:VD}
  L=\left( \ell_i \right)_{i=1,\ldots,\gamma}\, . 
\end{equation}
just corresponds to the nontrivial linear forms in  $Q(z)$.
Since we look for a $\nu \in {\cal V}$ with $Q(z)\nu=0$, 
we are interested in the sign patterns that $Lz$ can assume. 
So we define the set $\mathcal{L}^+$ of all
\emph{sign pattern matrices} $\Sigma\in \{-1,0,1\}^{s\times p}$
that are \emph{$L^+$-matrices} and that are realized by $Q(z)$ for some $z$.
Since the  mapping (\ref{map-vec}) associates a {\em signature vector}
$\sigma =\psi(\Sigma)\in \{-1,0,1\}^\gamma $ to $\Sigma\in \{-1,0,1\}^{s\times p}$
one arrives at
\begin{equation}\label{eq:LD}
  \begin{split}
    \mathcal{L}^+ 
    &:= \Bigl\{ \Sigma \in \{-1,0,1\}^{s\times p} \, \Big| \   \text{$\Sigma$ is an $L^+$-matrix},\\
    &\qquad   \exists z\, \in \R^n\; \text{with $\sigma_i\, \left(Lz\right)_i >0$ if
      $\sigma_i \neq 0$ and $\left(L\, z\right)_i =0$ if $\sigma_i = 0$}
    \Bigr\}\, .
  \end{split}
\end{equation}
\begin{fact}
  \label{fact:f1}
  Assume  $\mathcal{L}^+$ is
  nonempty and let $\Sigma\in\mathcal{L}^+$ and $\sigma=\psi(\Sigma)$. 
  Then there exists a vector
  $z\in\R^n$ with 
  \begin{displaymath}
    \sigma_i\, \left(L\, z\right)_i > 0\text{, if $\sigma_i\neq 0$},\ \ 
    \left(L\, z\right)_i = 0\text{, if $\sigma_i = 0$}
  \end{displaymath}
  so that $\sigma=\sign\brac{L\, z}$. Moreover, for each such
  $z\in\R^n$, there exists a positive $\nu=\nu(z)$ with $Q(z)\, \nu =
  0$ by Theorem~\ref{theo:L+-matrices}. By the discussion of (\ref{eq:Qz_condi}),
  this implies
  that the pair ($z$, $\nu$) satisfies $H(\nu)\, z = 0$.
\end{fact}

\vspace{2mm}\begin{theorem}
  \label{lem:main}
  Consider a dynamical system defined by a mass action network as
  described in Section~\ref{sec:dyn-sys-mass-action}. Recall the
  matrix $Q(z)$ defined in (\ref{eq:Teq}), the matrix $L$ defined
  in (\ref{map-vec})\&(\ref{eq:VD}) and the set $\mathcal{L}^+$ defined in
  (\ref{eq:LD}). If there exist an element $\Sigma=\sign\brac{Q(z)}\in\mathcal{L}^+$ 
  and an element $\omega\in\ker\brac{Z^T}$ with
  \begin{equation}
    \label{eq:li3}
    \sign(\omega) = \sign(z),
  \end{equation}
  then there exists a solution $\nu\in\Rp^p$ to $Q(z)\, \nu = 
  0$ and the Jacobian $G(\xi , \eta ,\nu)$ of the reduced system has
  zero as an eigenvalue with algebraic multiplicity $\geq 1$. The
  corresponding steady state in original ($k$, $x$)-coordinates is
  given by
  \begin{subequations}
    \begin{align}
      x & = \left(x_i\right)_{i=1,\ldots,n} , \\
      x_i &= \begin{cases}
        \frac{\omega_i}{z_i}\text{, if $z_i\neq 0$ ,} \\
        \bar x_i > 0\text{, arbitrary, if $z_i=0$ ,}
      \end{cases} \\
      k &= \diag\left(\phi\left(x^{-1}\right)\right)\, E\, \nu\, .
    \end{align}
    The corresponding $(\xi,\eta)$-coordinates  in $G(\xi , \eta ,\nu)$
    are then given by (\ref{eq:reduced-system0}).
  \end{subequations}
\end{theorem}
\begin{proof} The statements follow directly from
  Lemma~\ref{sec:lem_pos_x_sign_z} and Fact~\ref{fact:f1}.
\end{proof}

\vspace{2mm}The condition (\ref{eq:li3}) can be tested by examining
the following linear inequality systems defined by orthants of
$R^n$. To establish these we identify each orthant by its sign pattern
$\delta\in\{-1,0,1\}^n$: let $x\in\R^n$, then the sign pattern of $x$
is defined as $\delta:=\sign(x)$ and the orthant containing $x$ is
given by $\R_\delta^n := \{ x\in\R^n | \sign(x) = \delta\}$. To find
$z$ and $\omega$ satisfying (\ref{eq:li3}) for a given signature
$\sigma =\psi(\Sigma)$ for an $L^+$-matrix $\Sigma$ 
then amounts to finding an orthant $\R_\delta^n$ such that
\begin{subequations}
  \begin{gather}
    \label{eq:orthi_sys_1}
     \sigma_i\, \left(Lz\right)_i > 0\, \text{ if
      $\sigma_i\neq 0$},\;\ \ \left(Lz\right)_i = 0\, \text{ if $\sigma_i
      = 0$}, \\
    \label{eq:orthi_sys_2}
    Z^T\, \omega = 0,\;\  \text{with }\ \;
    \delta_i\, \omega_i >0,\, \delta_i\, z_i >0 \ \text{ if $\delta_i
      \neq 0$}\ \text{and} \  \omega_i = 0,\, z_i=0\ \text{ if $\delta_i=0$}.
  \end{gather}
\end{subequations}

\vspace{2mm}\begin{corollary}
  If there exists a signature $\Sigma\in\mathcal{L}^+$ and an orthant
  $\R_\delta^n$, such that the linear inequality system
  (\ref{eq:orthi_sys_1}), (\ref{eq:orthi_sys_2}) is feasible, then the
  reduced Jacobian  $G(\xi , \eta ,\nu)$ has zero as an eigenvalue with
  algebraic multiplicity $\geq 1$.
\end{corollary}

\vspace{2mm}\begin{remark}
  In the previous discussion we have only considered positive kernel
  vectors of $Q(z)$. However the set ${\cal V}$ can contain nonnegative
  vectors $\nu$. Thus, suppose $\nu$ contains the facet of
  $\Rnn^p$ given by $\bigl\{\nu\in\Rnn^p | \nu_i=0$,
  $\nu_j>0$, $i\neq j=1$, \ldots, $p\bigr\}$. Then one may fix
  $\nu_i=0$, replace $Q(z)$ in the discussion above by the
  submatrix $\tilde Q(z)$ obtained by deleting the $i$-th column (and
  eventually occurring zero rows) and obtain the remaining $\nu_i$
  by asking for positive kernel vectors of $\tilde Q(z)$ (i.e.\ by
  establishing the $L^+$-property for $\tilde Q(z)$).
\end{remark}

\vspace{2mm}\begin{remark}
  The condition (\ref{eq:orthi_sys_1})
  tests  whether the given  $L^+$-matrix $\Sigma$ belongs to $\mathcal{L}^+$.
  By
  the definition of the matrix $L$ this requires the labeling of the
   hyperplane arrangement given by $Lz$, which is computationally
    expensive (for an algorithm see \cite{arr-001}, 
   \cite{arr-003}). We have shown that all $z$ satisfying (\ref{eq:orthi_sys_1})
   lead to a positive null vector of $Q(z)$.
  The condition (\ref{eq:orthi_sys_2}) then stands for the compatibility 
  with the kernel of $Z^T$: It tests whether there is a $z$ in the solution set of 
  (\ref{eq:orthi_sys_1}) that possesses a signature thats is compatible with $\ker\brac{Z^T}$.
  Since one has to
  decide whether or not one of the systems (\ref{eq:orthi_sys_1}),
  (\ref{eq:orthi_sys_2}) is feasible the overall procedure can be
  computationally demanding, even though the individual steps only
  involve simple matrix computations.
\end{remark}

\vspace{3mm}\section{Saddle node bifurcations for the G1/S transition in budding
  yeast} 
\label{sec:application}\mbox{}

The networks displayed in (\ref{eq:net_ten}) and (\ref{eq:net_bin})
below are competing hypotheses describing the G1/S transition in
budding yeast. Both networks are biologically  plausible and hard to
distinguish experimentally \cite{fein-025}.
\begin{equation}
  \label{eq:net_ten}
  \scalebox{0.8}{ 
    \begin{minipage}{1.0\linewidth}
      \begin{displaymath}
        \begin{split}
          &\xymatrix{
            {\bf [Sic1P]} \ar [r] ^{\bf k_3} & {\bf [0]} \ar @<.4ex> @{-^>}
            [r] ^{\bf k_1} & \ar @{-^>} [l] ^{\bf k_2} {\bf [Sic1]} } \\
          &\xymatrix{
            {\bf [Clb] + [Sic1]} \ar @<.4ex> @{-^>} [r] ^{\bf k_4} & \ar
            @{-^>} [l] ^{\bf k_5} {\bf [Clb\cdot Sic1]} \ar [dr] ^{\bf k_6}
            & \\ 
            & & {\bf [Clb]} \\
            {\bf [Clb] + [Sic1P]} \ar @<.4ex> @{-^>} [r] ^{\bf k_7} & \ar
            @{-^>} [l] ^{\bf k_8} {\bf [Clb\cdot Sic1P]} \ar [ur] ^{\bf k_9}
            & } \\ 
          &\xymatrix{
            {\bf [Clb\cdot Sic1] + [Clb]} \ar @<.4ex> @{-^>} [r] ^{\bf k_{10}}
            & \ar @{-^>} [l] ^{\bf k_{11}} {\bf [Clb\cdot Sic1 \cdot Clb]} \ar
            [r] ^{\bf k_{12}} & {\bf [Clb\cdot Sic1P] + [Clb]}
          } \\
          &\xymatrix{
            {\bf [Sic1P] + [Cdc14]} \ar @<.4ex> @{-^>} [r] ^{\bf k_{13}} &
            \ar @{-^>} [l] ^{\bf k_{14}} {\bf [Sic1P\cdot Cdc14]} \ar [r]
            ^{\bf k_{15}} & {\bf [Sic1] + [Cdc14]} 
          } \\ 
          &\xymatrix{
            {\bf [Clb\cdot Sic1P] + [Cdc14]}\ar @<.4ex> @{-^>} [r] 
            ^{\bf k_{16}} & \ar @{-^>} [l] ^{\bf k_{17}} {\bf [Clb\cdot
              Sic1P\cdot Cdc14]} \ar [r] ^{\bf k_{18}} & {\bf [Clb\cdot Sic1]
              + [Cdc14]}
          }
        \end{split}
      \end{displaymath}
    \end{minipage}
  }
\end{equation}

\begin{equation}
  \label{eq:net_bin}
  \scalebox{0.8}{
    \begin{minipage}{1.0\linewidth}
      \begin{displaymath}
        \begin{split}
          &\xymatrix{
            {\bf [Sic1P]}\ar [r] ^{\bf k_3} & {\bf [0]} \ar  @<.4ex> @{-^>}[r]
            ^{\bf k_1} & {\bf [Sic1] } \ar @{-^>}[l]^{\bf k_2}& & 
          } \\
          &\xymatrix{
            {\bf [Sic1\cdot Clb]}\ar  @<.4ex> @{-^>}[r] ^{\bf k_{4}} \ar
            [ddr] ^{\bf k_9} & \ar @{-^>} ^{\bf k_5} [l] {\bf [Clb] +
              [Sic1]} \ar @<.4ex> @{-^>}[r] ^{\bf k_6} & \ar @{-^>} ^{\bf
              k_7} [l] {\bf [Clb\cdot Sic1]} \ar [dr] ^{\bf k_8} & \\ 
            & & & {\bf [Clb]} \\
            & {\bf [Clb] + [Sic1P]} \ar  @<.4ex> @{-^>}
            [r] ^{\bf k_{10}} & \ar 
            @{-^>}[l] ^{\bf k_{11}} {\bf [Clb\cdot Sic1P]} \ar [ur]
            ^{\bf k_{12}} & 
          }\\
          &\xymatrix{
            {\bf [Sic1P] + [Cdc14]} \ar @<.4ex> @{-^>} [r] ^{\bf k_{13}} &
            \ar @{-^>} [l] ^{\bf k_{14}} {\bf [Sic1P\cdot Cdc14]} \ar [r]
            ^{\bf k_{15}} & {\bf [Sic1] +[Cdc14]} & 
          } \\ 
          &\xymatrix{
            {\bf [Clb\cdot Sic1P] + [Cdc14]} \ar @<.4ex> @{-^>} [r] ^{\bf
              k_{16}} &  \ar @{-^>} [l] ^{\bf k_{17}} {\bf [Clb \cdot
              Sic1P\cdot Cdc14]}  \ar [r] ^{\bf k_{18}} & {\bf [Clb\cdot
              Sic1] + [Cdc14]}
          }
        \end{split}
      \end{displaymath}
    \end{minipage}
  }
\end{equation}

Switching is a desired property of models describing the G1/S
transition: depending on its past a trajectory should move to
different regions of state space, associated with the G1 and the S
phase of cell cycle. Classically this has been realized by choosing
rate constants and total concentrations, such that the ODE system
shows bistability and hence hysteretic behaviour
\cite{cyc-005,sig-042}. For example in \cite{fein-025},
multistationarity has been established for both models, indicating
that both may be valid models. Here we demonstrate the applicability
of our results by confirming switching for both networks. We show that
both models satisfy the conditions of Theorem~\ref{lem:main} and
compute states and rate constants where the Jacobian has a defective
eigenvalue. We verify by numerical continuation that the system
undergoes a saddle-node bifurcation, as generically expected, so that
the codimension-$1$ stable manifold of the saddle-node and - after
bifurcation - the one of the saddle represents a switching surface.

For the network given in \eqref{eq:net_ten} one obtains using the
stoichiometric matrix $S$ given in Appendix~\ref{sec:data-tern}:

\begin{flushleft}
  \scalebox{0.5}{
    \begin{minipage}{1.0\linewidth}
      \begin{displaymath}
        Q(z) = \left[
          \begin{array}{ccccccccccc}
            -z_1 & 0 & 0 & 0 & 0 & 0 & -z_9 & -z_4 & -z_9 & z_7-z_9 &
            z_8-z_9 \\
            0 & 0 & 0 & 0 & 0 & 0 & z_2-z_9 & 0 & z_5-z_9 & z_7-z_9 &
            z_8-z_9 \\ 
            0 & z_1+z_3-z_4 & 0 & 0 & 0 & 0 & z_1+z_3-z_9 & z_1+z_3-z_4 &
            z_1+z_3-z_9 & z_1+z_3-z_9 & z_8-z_9 \\
            0 & 0 & z_2+z_3-z_5 & 0 & 0 & 0 & -z_5+z_9 & 0 & -z_5+z_9 &
            -z_5+z_9 & -z_8+z_9 \\
            0 & 0 & 0 & z_3+z_4-z_9 & 0 & 0 & z_3+z_4-z_9 & 0 & z_3+z_4-z_9
            & z_3+z_4-z_9 & z_3+z_4-z_9 \\
            0 & 0 & 0 & 0 & z_2+z_6-z_7 & 0 & 0 & 0 & 0 & z_2+z_6-z_7 & 0 \\
            0 & 0 & 0 & 0 & 0 & z_5+z_6-z_8 & 0 & 0 & 0 & 0 & z_5+z_6-z_8
          \end{array}
        \right]
      \end{displaymath}
    \end{minipage}
  }
\end{flushleft}

\begin{figure}
  \centering
  \subfigure[Continuation for network \eqref{eq:net_ten}]{\includegraphics[width=.4\linewidth]{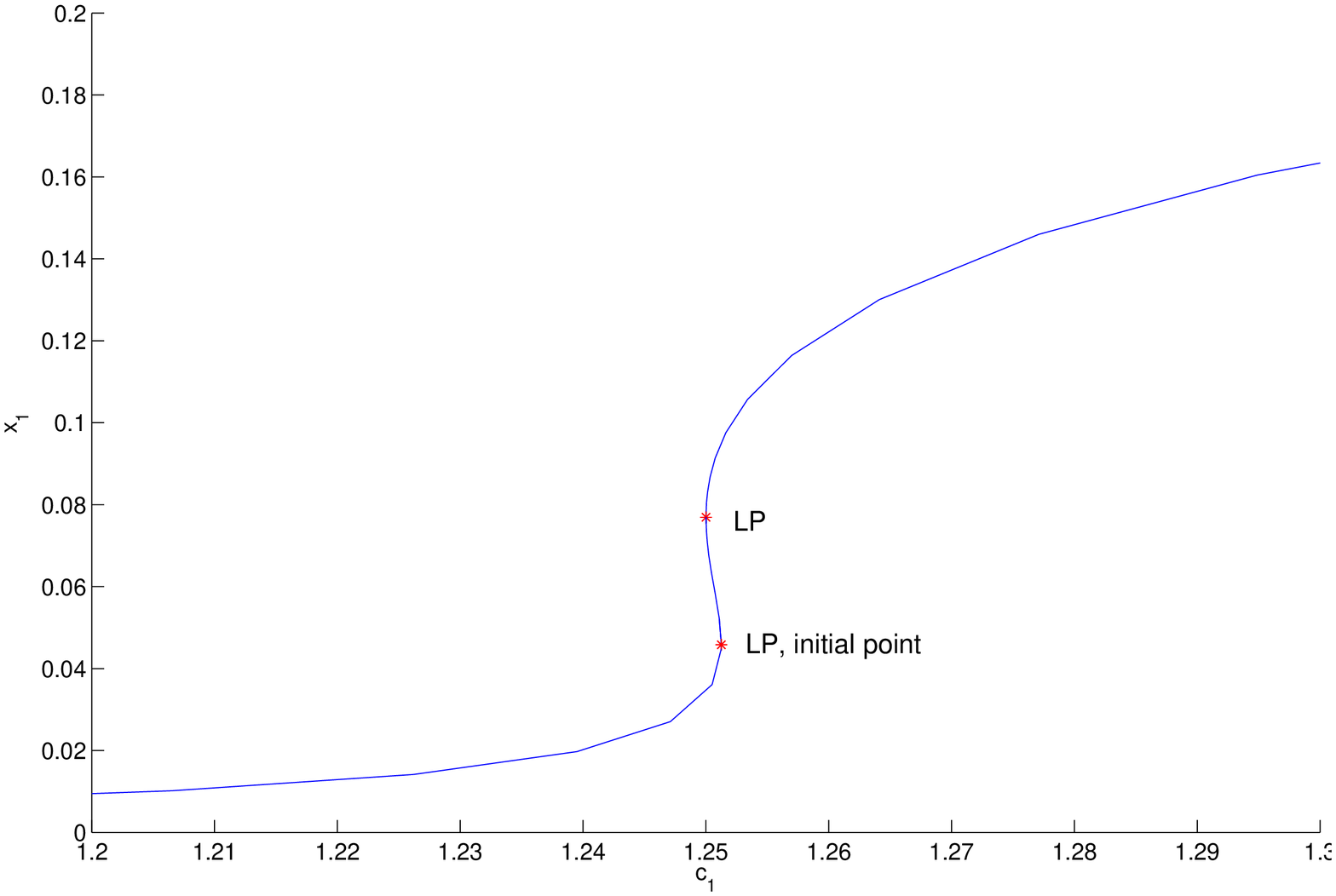}}
  \subfigure[Continuation for network \eqref{eq:net_bin}]{\includegraphics[width=.4\linewidth]{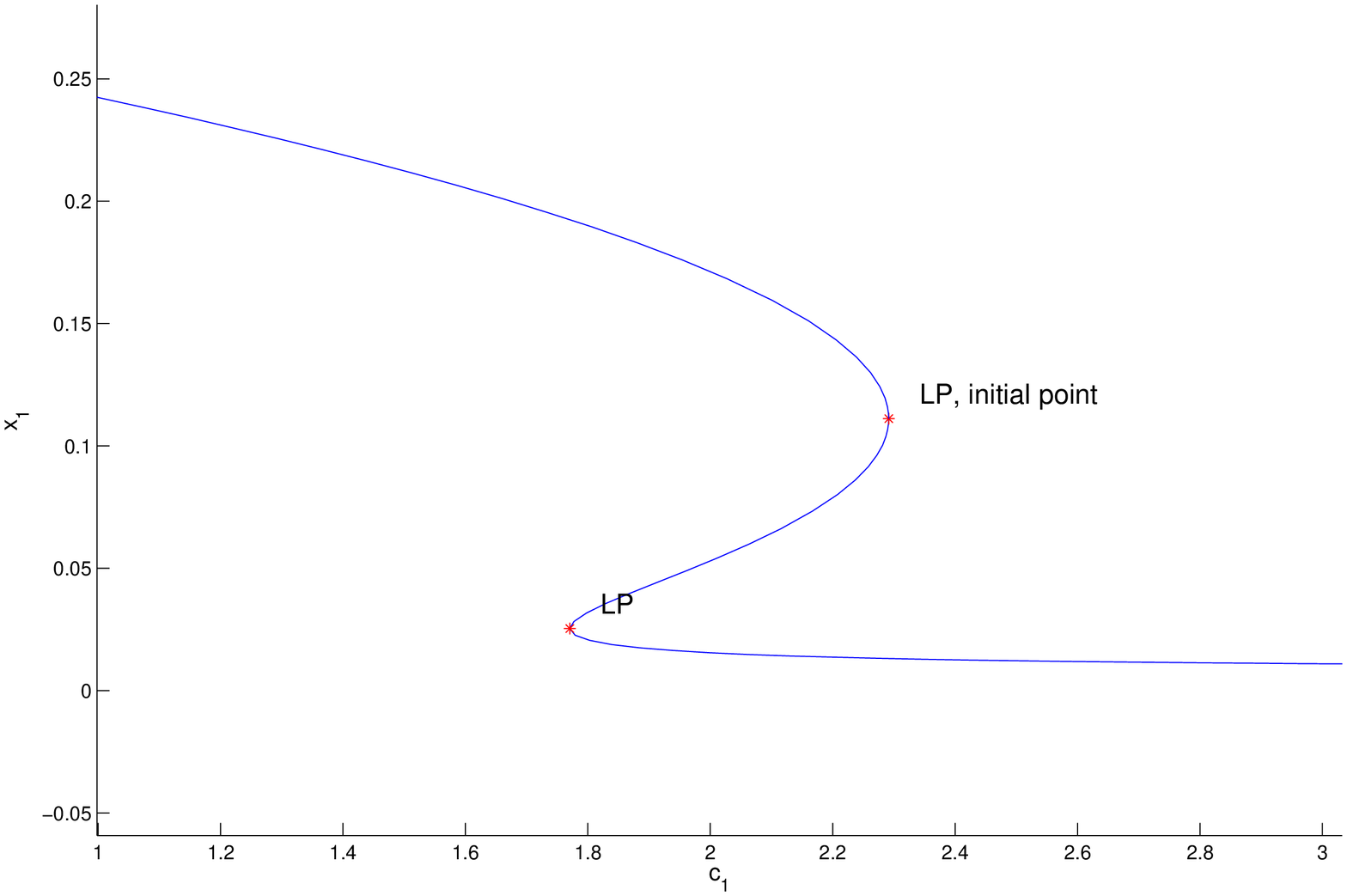}}
  \caption{\label{fig:conti_tc} Numerical continuation for the
    networks \eqref{eq:net_ten} and \eqref{eq:net_bin} using rate
    constants $\tilde k$ and initial condition $\tilde x$ given in
    (\ref{eq:x_tern}), (\ref{eq:k_tern}) for network
    (\ref{eq:net_ten}) and in (\ref{eq:x_bin}), (\ref{eq:k_bin}) for
    network (\ref{eq:net_bin}). In both cases the upper and lower branches
    correspond to exponentially stable steady states. The total
    concentration $c_1$ is used as a bifurcation parameter.}
\end{figure}

\vspace{2mm}\noindent From the last three rows of $Q(z)$ one has that
positive $\nu$ with $Q(z)\, \nu=0$ exist only if 
\begin{equation}
\label{eq:eq_const_tern}
  z_3+z_4-z_9 = 0,\; z_2+z_6-z_7 =0,\; z_5+z_6-z_8 =0
\end{equation}
and hence, for example,
\begin{displaymath}
  z_2 = -z_6+z_7,\; z_3 = -z_4 + z_9,\; z_5 = -z_6+z_8\ .
\end{displaymath}
In this case colum 4, 5 and 6 will be the zero column, indicating that
$\nu_4$, $\nu_5$, $\nu_6>0$ are unconstrained. Thus we
need only consider the matrix
\begin{displaymath}
  Q_s(z) = 
  \left[
    \begin{array}{cccccccc}
      -\pi_{8}\, z & 0 & 0 & -\pi_{10}\, z & -\pi_{9}\, z & -\pi_{10}\, z
      & \pi_{6}\, z & \pi_{5}\, z \\ 
      0 & 0 & 0 & \pi_{2}\, z & 0 & -\pi_{4}\, z & \pi_{6}\, z &
      \pi_{5}\, z \\ 
      0 & \pi_{3}\, z & 0 & \pi_{7}\, z & \pi_{3}\, z & \pi_{7}\, z &
      \pi_{7}\, z & \pi_{5}\, z \\ 
      0 & 0 & \pi_{1}\, z & \pi_{4}\, z & 0 & \pi_{4}\, z & \pi_{4}\, z &
      -\pi_{5}\, z
    \end{array}
  \right]
\end{displaymath}
with
\begin{align*}
 \pi_{1}\, z &:= -z_{4}+z_{7}-z_{8}+z_{9} &
 \pi_{2}\, z &:= z_{5}+z_{7}-z_{8}-z_{9} \\
 \pi_{3}\, z &:= z_{1}-2 z_{4}+z_{9} &
 \pi_{4}\, z &:= -z_{5}+z_{9} \\
 \pi_{5}\, z &:= z_{8}-z_{9} &
 \pi_{6}\, z &:= z_{7}-z_{9} \\
 \pi_{7}\, z &:= z_{1}-z_{4} &
 \pi_{8}\, z &:= z_{1} \\
 \pi_{9}\, z &:= z_{4} &
 \pi_{10}\, z &:= z_{9}
\end{align*}
For example the system
\begin{equation}
  \label{eq:ineq_const_tern}
  \begin{split}
    \pi_{1}\, z < 0,\; \pi_{2}\, z >0,\; \pi_{3}\, z <0,\; \pi_{4}\, z >0,\;
    \pi_{5}\, z <0,\\ \pi_{6}\, z >0,\;  \pi_{7}\, z > 0,\; \pi_{8}\, z >
    0,\; \pi_{9}\, z > 0,\; \pi_{10}\, z < 0 \\
    z_1>0,\; z_2>0,\; z_3<0,\; z_4>0,\; z_5<0,\; z_6>0,\; z_7>0,\;
    z_8<0,\; z_9<0 \\
    \omega_1>0,\; \omega_2>0,\; \omega_3<0,\; \omega_4>0,\;
    \omega_5<0,\; \omega_6>0,\; \omega_7>0,\; \omega_8<0,\; \omega_9<0
  \end{split}
\end{equation}
is feasible.  Let $z\in\R^9$ such that \eqref{eq:eq_const_tern} and
  \eqref{eq:ineq_const_tern} hold. Then 
\begin{displaymath}
  \sign\brac{Q_s(z)} = \left[
    \begin{array}{rrrrrrrr}
      -1&0&0&1&-1&1&1&-1 \\
      0&0&0&1&0&-1&1&-1 \\
      0&-1&0&1&-1&1&1&-1 \\
      0&0&-1&1&0&1&1&1
    \end{array}
  \right]
\end{displaymath}
is an $L^+$-matrix (cf. (c,d) of Theorem~\ref{theo:L+-matrices}). 
One obtains, for example, the feasible points
\begin{displaymath}
  \tilde z=\left(13,\, 2,\, -10,\, 8,\, -6,\, 2,\, 4,\, -4,\, -2
  \right)^T,\;
  \tilde \omega = \left(1,\, 1,\, -1,\, 6,\, -1,\, 1,\, 1,\, -2,\, -1
  \right)^T.
\end{displaymath}
Vectors $\tilde z$ and $\tilde \omega$ yield the state vector
\begin{equation}
  \label{eq:x_tern}
  \tilde x = \frac{\tilde\omega}{\tilde z} = \left(
    \frac{1}{13},\, 
    \frac{1}{2},\, 
    \frac{1}{10},\, 
    \frac{3}{4},\, 
    \frac{1}{6},\, 
    \frac{1}{2},\, 
    \frac{1}{4},\, 
    \frac{1}{2},\, 
    \frac{1}{2}
  \right)
\end{equation}
For the matrix $Q(z)$ evaluated at $\tilde z$ one has
\begin{displaymath}
  Q(\tilde z) = \left[
    \begin{array}{rrrrrrrrrrr}
      -13&0&0&0&0&0&2&-8&2&6&-2 \\
      0&0&0&0&0&0&4&0&-4&6&-2 \\
      0&-5&0&0&0&0&5&-5&5&5&-2 \\
      0&0&-2&0&0&0&4&0&4&4&2 \\
    \end{array}
  \right].
\end{displaymath}
The kernel of $Q(\tilde z)$ contains the following positive vector
\begin{displaymath}
  \tilde\nu = \left(
    18,\, 
    1,\, 
    2000,\, 
    1,\, 
    1,\, 
    1,\, 
    1,\, 
    \frac{246}{35},\, 
    \frac{5114}{105},\, 
    \frac{1996}{5},\, 
    \frac{23146}{21}
  \right)^T.
\end{displaymath}
Vectors $\tilde x$ and $\tilde \nu$ yield the rate constants
\begin{equation}
  \label{eq:k_tern}
  \begin{split}
    \tilde k &= \Biggl(
    \frac{1121}{15},\, 
    234,\, 
    2,\, 
    \frac{178204}{3},\, 
    \frac{4}{3},\, 
    \frac{328}{35},\, 
    40000,\, 
    \frac{72006}{5},\, 
    \frac{10228}{35},\, 
    \frac{1303760}{63},\\
    &\qquad 
    2,\, 
    \frac{65146}{21},\, 
    \frac{8004}{5},\, 
    4,\, 
    \frac{7984}{5},\, 
    \frac{92668}{7},\, 
    2,\, 
    \frac{46292}{21}
    \Biggr)^T.
  \end{split}
\end{equation}
A numerical continuation with this $\tilde k$ and initial
condition $\tilde x$ verifies that at the dynamical system undergoes a
saddle-node bifurcation at ($\tilde k$, $\tilde x$), cf.\
Fig.~\ref{fig:conti_tc}(a). 

\vspace{2mm}
For the network given in \eqref{eq:net_bin} one obtains using the
stoichiometric matrix $S$ as given in Appendix~\ref{sec:data-bin}:
\begin{flushleft}
  \scalebox{0.5}{
    \begin{minipage}{1.0\linewidth}
      \begin{displaymath}
        Q(z) = \left[
          \begin{array}{cccccccccccc}
            -z_{1} & 0 & 0 & 0 & 0 & 0 & -z_{4} & -z_{9} & z_{7}-z_{9}
            & -z_{9} & -z_{4}+z_{8}-z_{9} & z_{8}-z_{9}      \\ 
            0 & 0 & 0 & 0 & 0 & 0 & 0 & z_{2}-z_{9} & z_{7}-z_{9} &
            z_{5}-z_{9} & z_{8}-z_{9} & z_{8}-z_{9} \\ 
            0 & -z_{1}-z_{3}+z_{9} & 0 & 0 & 0 & 0 & 0 &
            -z_{1}-z_{3}+z_{9} & -z_{1}-z_{3}+z_{9} &
            -z_{1}-z_{3}+z_{9} & -z_{1}-z_{3}+z_{9} &
            -z_{1}-z_{3}+z_{9} \\ 
            0 & 0 & z_{1}+z_{3}-z_{4} & 0 & 0 & 0 & z_{1}+z_{3}-z_{4}
            & 0 & 0 & 0 & -z_{4}+z_{8} & -z_{4}+z_{8} \\ 0 & 0 & 0 &
            z_{2}+z_{3}-z_{5} & 0 & 0 & 0 & 0 & 0 & z_{2}+z_{3}-z_{5}
            & z_{2}+z_{3}-z_{8} & z_{2}+z_{3}-z_{8} \\ 
            0 & 0 & 0 & 0 & z_{2}+z_{6}-z_{7} & 0 & 0 & 0 &
            z_{2}+z_{6}-z_{7} & 0 & 0 & 0 \\ 
            0 & 0 & 0 & 0 & 0 & z_{5}+z_{6}-z_{8} & 0 & 0 & 0 & 0 &
            z_{5}+z_{6}-z_{8} & z_{5}+z_{6}-z_{8}
          \end{array}
        \right]
      \end{displaymath}
    \end{minipage}
  }
\end{flushleft}

\vspace{2mm}\noindent From rows 3, 6 and 7 of $Q(z)$ one has that
positive $\nu$ with $Q(z)\, \nu=0$ exist only if 
\begin{equation}
  \label{eq:eq_const_bin}
  -z_{1}-z_{3}+z_{9}=0,\; z_{5}+z_{6}-z_{8}=0,\; z_{2}+z_{6}-z_{7}=0
\end{equation}
and hence, for example,
\begin{displaymath}
  z_1 = -z_3+z_9,\; z_2 = z_5 -z_8 + z_7,\; z_6= -z_5 + z_8\ .
\end{displaymath}
In this case colum 2, 5 and 6 will be the zero column, indicating that
$\nu_2$, $\nu_5$, $\nu_6>0$ are unconstrained. Thus we need only
consider the matrix
\begin{displaymath}
  Q_s(z) = 
  \left[
    \begin{array}{ccccccccc}
      \pi_{10}\, z & 0 & 0 & -\pi_{11}\, z & -\pi_{12}\, z & \pi_{7}\,
      z & -\pi_{12}\, z & \pi_{4}\, z & \pi_{9}\, z \\ 
      0 & 0 & 0 & 0 & \pi_{3}\, z & \pi_{7}\, z & \pi_{8}\, z &
      \pi_{9}\, z & \pi_{9}\, z \\ 
      0 & \pi_{5}\, z & 0 & \pi_{5}\, z & 0 & 0 & 0 & \pi_{6}\, z &
      \pi_{6}\, z \\ 
      0 & 0 & \pi_{2}\, z & 0 & 0 & 0 & \pi_{2}\, z & \pi_{1}\, z &
      \pi_{1}\, z
    \end{array}
  \right]
\end{displaymath}
with
\begin{align*}
  \pi_{1}\, z &= z_{3}+z_{5}+z_{7}-2 z_{8} &
  \pi_{2}\, z &= z_{3}+z_{7}-z_{8} \\
  \pi_{3}\, z &= z_{5}+z_{7}-z_{8}-z_{9} &
  \pi_{4}\, z &= -z_{4}+z_{8}-z_{9} \\
  \pi_{5}\, z &= -z_{4}+z_{9} &
  \pi_{6}\, z &= -z_{4}+z_{8} \\
  \pi_{7}\, z &= z_{7}-z_{9} &
  \pi_{8}\, z &= z_{5}-z_{9} \\
  \pi_{9}\, z &= z_{8}-z_{9} &
  \pi_{10}\, z &= z_{3}-z_{9} \\
  \pi_{11}\, z &= z_{4} &
  \pi_{12}\, z &= z_{9}
\end{align*}
One obtains the feasible inequality system
\begin{equation}
  \label{eq:ineq_const_bin}
  \begin{split}
    \pi_{1}\, z > 0,\;
    \pi_{2}\, z < 0,\; 
    \pi_{3}\, z > 0,\;  
    \pi_{4}\, z = 0,\; 
    \pi_{5}\, z > 0,\; 
    \pi_{6}\, z < 0,\; 
    \pi_{7}\, z > 0,\ \\
    \pi_{8}\, z > 0,\;
    \pi_{9}\, z < 0,\; 
    \pi_{10}\, z < 0,\; 
    \pi_{11}\, z < 0,\;
    \pi_{12}\, z < 0, \\
    z_1 > 0,\; 
    z_2 > 0,\;
    z_3 < 0,\;
    z_4 < 0,\;
    z_5 > 0,\;
    z_6 < 0,\;
    z_7 > 0,\;
    z_8 < 0,\; 
    z_9 < 0, \\
    \omega_1 > 0,\; 
    \omega_2 > 0,\;
    \omega_3 < 0,\;
    \omega_4 < 0,\;
    \omega_5 > 0,\;
    \omega_6 < 0,\;
    \omega_7 > 0,\;
    \omega_8 < 0,\; 
    \omega_9 < 0,
  \end{split}
\end{equation}
where $Q(z)$ is an $L^+$-matrix. For example the feasible points
\begin{displaymath}
  \tilde z = \left(9,\, 
    9,\,
    -11,\,
    -4,\,
    2,\,
    -8,\,
    1,\,
    -6,\,
    -2 \right)^T,\; 
  \tilde \omega = \left(
    1,\,
    1,\,
    -1,\,
    -1,\,
    4,\,
    -1,\,
    2,\,
    -1,\,
    -1
  \right)^T
\end{displaymath}
define the state vector
\begin{equation}
  \label{eq:x_bin}
  \tilde x = \left(
      \frac{1}{9},\, 
      \frac{1}{9},\, 
      \frac{1}{11},\, 
      \frac{1}{4},\, 
      2,\, 
      \frac{1}{8},\, 
      2,\, 
      \frac{1}{6},\,
      \frac{1}{2}
  \right)
\end{equation}
For the matrix $Q(z)$ evaluated at $\tilde z$ one obtains
\begin{displaymath}
  Q\left(\tilde z\right) = \left[
    \begin{array}{rrrrrrrrrrrr}
      -9 & 0 & 0 & 0 & 0 & 0 & 4 & 2 & 3 & 2 & 0 & -4 \\
      0 & 0 & 0 & 0 & 0 & 0 & 0 & 11 & 3 & 4 & -4 & -4 \\
      0 & 0 & 2 & 0 & 0 & 0 & 2 & 0 & 0 & 0 & -2 & -2 \\
      0 & 0 & 0 & -4 & 0 & 0 & 0 & 0 & 0 & -4 & 4 & 4
    \end{array}
    \right],
\end{displaymath}
with the positive kernel vector
\begin{displaymath}
  \tilde \nu = \left(
    297,\,
    11,\,
    22,\,
    11,\,
    11,\,
    11,\,
    440,\,
    1,\,
    11,\,
    451,\,
    456,\,
    6
  \right)^T.
\end{displaymath}
Finally one obtains for the rate constants
\begin{equation}
\label{eq:k_bin}
  \begin{split}
    \tilde k &= \bigl(
      1645,\,
      2673,\,
      9,\,
      22,\,
      92664,\,
      45738,\,
      112,\,
      3584,\,
      1850,\,
      91476,\\
      &\qquad 
      \frac{11}{2},\,
      \frac{451}{2},\,
      1584,\,
      \frac{11}{2},\,
      \frac{11}{2},\,
      1892,\,
      66,\,
      2772
    \bigr)^T.
  \end{split}
\end{equation}
Again, numerical continuation show a saddle node bifurcation at
($\tilde k$, $\tilde x$), cf.\ Fig.\ref{fig:conti_tc}(b).

\newpage

\appendix{}



\section{Saddle-node bifurcations in mass action networks}
\label{sec:saddle-node-mass-action}

This subsection is a recollection of some remarks concerning
saddle-node bifurcations in mass action system that were originally
made in \cite{fein-032}. We repeat them here to demonstrate the tight
connection between zero eigenvalues of $G(\xi , \eta ,\lambda )$ obtained via
(\ref{eq:bif_1})\&(\ref{eq:bif_2}) and saddle node bifurcations.

The following well-known theorem gives necessary and sufficient
conditions for a saddle-node bifurcation of the system $\dot{\xi} =
g\left(\xi,\eta,k\right)$ as defined in \eqref{eq:reduced-system}. Let $\mu$ be
the component of the parameter vector $\nu =(\eta,k)$ that will be used
as bifurcation parameter. 

\begin{theorem}[see e.g. 
\cite{ode-001}, p. 497]
  \label{theo:Saddle-node-bifurcation}
  Suppose $(\xi_0,\nu_0)$ is a zero of $g$ and suppose that the
  $s\times s$ matrix $G(\xi_0,\nu_0)=D_{\xi}\, g(\xi_0,\nu_0)$ has an algebraically
  simple eigenvalue $0$ with right eigenvector $b$ and left
  eigenvector $\beta ^T$. Furthermore suppose that the following
  conditions are satisfied:
  \begin{equation}
    \label{eq:saddle-node-condition}
    \beta ^T\, D_{\mu}\, g(\xi_0,\nu_0)\neq 0, \quad 
    \beta ^T\left[D_{\xi}^2\, g(\xi_0,\nu_0)(b,b)\right]\neq 0.
  \end{equation}
  Then there is a smooth curve of zeroes of $g$ passing through
  $(\xi_0,\nu_0)$. Depending on the signs of the expressions in
  \eqref{eq:saddle-node-condition}, there are no or two zeroes near
  $\xi_0$ for $\mu \neq \mu_0$ when the other components of $\nu_0$
  remain fixed. 
\end{theorem}

In the remainder of this section we examine in terms of the Jacobian
$J(\nu,x)$  when the conditions 
\eqref{eq:saddle-node-condition} are satisfied for the reduced system
\eqref{eq:reduced-system}.  We recall the relations 
(\ref{eq:def_LAM-1}),
(\ref{eq:def_k_lambda}) and 
(\ref{eq:reduced-system}). 
First observe that
Fact~\ref{fact:zero-eigenvalues-reduced} states conditions
guaranteeing that $G(\xi , \eta ,\lambda )$ has zero as an eigenvalue.
For general mass action networks its algebraic multiplicity is
expected to be $1$.  

Now, if $G(\xi_0,\eta_0, \nu_0)$ has an algebraically simple 
eigenvalue $0$, we add two comments regarding (\ref{eq:saddle-node-condition}) 
(cf. \cite{fein-032}):
\begin{enumerate}
\item[(N1)] Recall the monomial function $\phi(x)$,
  cf. Section~\ref{sec:dyn-sys-mass-action}. If $\mu$ is any rate
  constant $k_i$, then we have
  \begin{displaymath}
    D_{\mu}\, g(\xi_0,\mu_0) =  
    U^T\, S\, \diag\left(\phi(x_0)\right)\, e_i
  \end{displaymath}
  and therefore 
  \begin{displaymath}
    \beta ^T D_{\mu}\, g(\xi_0,\mu_0) = \phi_i\left(x_0\right)\beta 
    ^T\, U^T\, S\, e_i \neq 0
  \end{displaymath}
  for at least one $i$ (as $\phi_i(x_0)>0$ and $[U]=\im\left( S
  \right)$).
\item[(N2)]
  From the above Lemma~\ref{lem:zero_eigen_defective} we deduce
  \begin{equation}
    \label{eq:bifcond2}
    \begin{split}
      \beta^T\, \left[D_{\xi}^2\, g(\xi_0,\eta_0,k_0)(b,b)\right] \neq
      0 \Leftrightarrow  \\
      \alpha^T\, \left[D_{x}^2\, f(x(\xi_0,\eta_0),k_0)(A\,
        a,A\, a)\right]\neq 0 
    \end{split}
  \end{equation}
  with $A:=J(\lambda,x)$ and with left and right principal vectors $\alpha ^T$
  and $a$ of $J(\lambda,x)$. 
\end{enumerate}
As a consequence of this discussion, in particular of the comment (N1),  we obtained in \cite{fein-032} the
following remark concerning the originally given system 
\eqref{eq:system_xdot}, \eqref{eq:system_con_rel}:
\begin{remark}
  \label{cor:saddle_node_bio_chem_rn}
  The system \eqref{eq:system_xdot}, \eqref{eq:system_con_rel} has a
  saddle-node bifurcation at $(k_0,x_0)$ (within the plane $Z^T\, x =
  Z^T\, x_0 =: c$) if the following conditions are satisfied:
  \begin{enumerate}[{(}a{)}]
  \item\label{item:cond-a} $0$ is a defective eigenvalue of $J(\lambda_0,x_0)$  
  with $m_{alg}=m_{geo}+1$ 
  and the remaining eigenvalues have negative real parts.
  \item\label{item:cond-b} 
  $\alpha ^T\left[D_{x}^2\, f(x_0,k_0)(Aa,Aa)\right]\neq 0$
  is satisfied for left and right principal vectors $w_0^T$ and $v_0$
  of $J(\lambda_0,x_0)$.
  \end{enumerate}
\end{remark}

\newpage

\section{The data for network \eqref{eq:net_ten}}
\label{sec:data-tern}

\subsection{Species and complexes of network \eqref{eq:net_ten}}

\begin{displaymath}
  \begin{array}{|c|c||c|c|}\hline
   \text{Species} & x_i & \text{Complex} & y_i \\ \hline
   Sic1 & x_1 & 0 & y_1 \\
   Sic1P& x_2 & Sic1 & y_2 \\
   Clb & x_3 & Sic1P & y_3 \\
   Clb\cdot Sic1 & x_4 & Clb+Sic1 & y_4 \\
   Clb\cdot Sic1P & x_5 & Clb\cdot Sic1 & y_5 \\
   Cdc14 &  x_6 & Clb & y_6 \\
   Sic1P\cdot Cdc14 & x_7 & Clb + Sic1P & y_7 \\ 
   Clb\cdot Sic1P\cdot Cdc14 & x_8 & Clb\cdot Sic1P & y_8 \\
   Clb\cdot Sic1 \cdot Clb & x_9 & Clb\cdot Sic1 + Clb & y_9 \\
   & & Clb\cdot Sic1\cdot Clb & y_{10} \\
   & & Clb\cdot Sic1P + Clb & y_{11} \\
   & & Sic1P + Cdc14 & y_{12} \\
   & & Sic1P\cdot Cdc14 & y_{13} \\
   & & Sic1 + Cdc14 & y_{14} \\
   & & Clb\cdot Sic1P + Cdc14 & y_{15} \\
   & & Clb\cdot Sic1P \cdot Cdc14 & y_{16} \\
   & & Clb\cdot Sic1 + Cdc14 & y_{17} \\ \hline
  \end{array}
\end{displaymath}

\subsection{Ordinary differential equations}

\begin{subequations}
  \begin{align*}
    \dot x_1 &= k_{1}-k_{2}\, x_{1}-k_{4}\, x_{1}\, x_{3} + k_{5}\,
    x_{4} + k_{15}\, x_{7} \\
    \dot x_2 &= -k_{3}\, x_{2}-k_{7}\, x_{2}\, x_{3} + k_{8}\,
    x_{5} -k_{13}\, x_{2}\, x_{6} + k_{14}\, x_{7}\\ 
    \dot x_3 &= -k_{4}\, x_{1}\, x_{3} + k_{5}\, x_{4} + k_{6}\,
    x_{4}-k_{7}\, x_{2}\, x_{3} + k_{8}\, x_{5} \\
    &\qquad + k_{9}\,
    x_{5}-k_{10}\, x_{3}\, x_{4} + k_{11}\, x_{9} + k_{12}\, x_{9}\\ 
    \dot x_4 &= k_{4}\, x_{1}\, x_{3}-k_{5}\, x_{4}-k_{6}\,
    x_{4}-k_{10}\, x_{3}\, x_{4} + k_{11}\, x_{9} + k_{18}\, x_{8}\\ 
    \dot x_5 &= k_{7}\, x_{2}\, x_{3}-k_{8}\, x_{5}-k_{9}\, x_{5} +
    k_{12}\, x_{9}-k_{16}\, x_{5}\, x_{6} + k_{17}\, x_{8}\\ 
    \dot x_6 &= -k_{13}\, x_{2}\, x_{6} + k_{14}\, x_{7} + k_{15}\,
    x_{7}-k_{16}\, x_{5}\, x_{6} + k_{17}\, x_{8} + k_{18}\, x_{8}\\ 
    \dot x_7 &= k_{13}\, x_{2}\, x_{6}-k_{14}\, x_{7}-k_{15}\, x_{7}\\
    \dot x_8 &= k_{16}\, x_{5}\, x_{6}-k_{17}\, x_{8}-k_{18}\, x_{8}\\
    \dot x_9 &= k_{10}\, x_{3}\, x_{4}-k_{11}\, x_{9}-k_{12}\, x_{9}
  \end{align*}
\end{subequations}

\subsection{Conservation relations}
\begin{subequations}
  \begin{align*}
Z_1^Tx= &  \ \ x_{6} + x_{7} + x_{8} \ \ = c_1 \\
Z_2^Tx=  &   x_{3} + x_{4} + x_{5} + x_{8} + 2\, x_{9} = c_2
   \end{align*}
\end{subequations}

\subsection{The stoichiometric matrix:}

\begin{center}
\scalebox{0.5}{
  \begin{minipage}{1.0\linewidth}
    \begin{displaymath}
      S = 
      \left[
        \begin{array}{rrrrrrrrrrrrrrrrrr}
          1 & -1 & 0 & -1 & 1 & 0 & 0 & 0 & 0 & 0 & 0 & 0 & 0 & 0 & 1 & 0 & 0 & 0 \\
          0 & 0 & -1 & 0 & 0 & 0 & -1 & 1 & 0 & 0 & 0 & 0 & -1 & 1 & 0 & 0 & 0 & 0 \\
          0 & 0 & 0 & -1 & 1 & 1 & -1 & 1 & 1 & -1 & 1 & 1 & 0 & 0 & 0 & 0 & 0 & 0 \\
          0 & 0 & 0 & 1 & -1 & -1 & 0 & 0 & 0 & -1 & 1 & 0 & 0 & 0 & 0 & 0 & 0 & 1 \\
          0 & 0 & 0 & 0 & 0 & 0 & 1 & -1 & -1 & 0 & 0 & 1 & 0 & 0 & 0 & -1 & 1 & 0 \\
          0 & 0 & 0 & 0 & 0 & 0 & 0 & 0 & 0 & 0 & 0 & 0 & -1 & 1 & 1 & -1 & 1 & 1 \\
          0 & 0 & 0 & 0 & 0 & 0 & 0 & 0 & 0 & 0 & 0 & 0 & 1 & -1 & -1 & 0 & 0 & 0 \\
          0 & 0 & 0 & 0 & 0 & 0 & 0 & 0 & 0 & 0 & 0 & 0 & 0 & 0 & 0 & 1 & -1 & -1 \\
          0 & 0 & 0 & 0 & 0 & 0 & 0 & 0 & 0 & 1 & -1 & -1 & 0 & 0 & 0 & 0 & 0 & 0
        \end{array}
      \right]
    \end{displaymath}
  \end{minipage}
}
\end{center}

\subsection{The vector of reaction rates:}

\begin{displaymath}
  \begin{split}
    v(k,x) &= \bigl(
      k_{1},\, 
      k_{2} x_{1},\, 
      k_{3} x_{2},\, 
      k_{4} x_{1} x_{3},\, 
      k_{5} x_{4},\, 
      k_{6} x_{4},\, 
      k_{7} x_{2} x_{3},\, 
      k_{8} x_{5},\, 
      k_{9} x_{5},\, 
      k_{10} x_{3} x_{4},\\
      &\qquad 
      k_{11} x_{9},\, 
      k_{12} x_{9},\, 
      k_{13} x_{2} x_{6},\, 
      k_{14} x_{7},\, 
      k_{15} x_{7},\, 
      k_{16} x_{5} x_{6},\, 
      k_{17} x_{8},\, 
      k_{18} x_{8}
    \bigr)^T
  \end{split}
\end{displaymath}

\section{The data for network \eqref{eq:net_bin}}
\label{sec:data-bin}

\subsection{Species and complexes of network \eqref{eq:net_bin}}
\label{sec:Variables-network-bin_com}

\begin{displaymath}
  \begin{array}{|c|c||c|c|} \hline
    \text{Species} & x_i & \text{Complex} & y_i \\ \hline
    Sic1 & x_1  & 0 & y_1 \\
    Sic1P & x_2 & Sic1 & y_2 \\
    Clb & x_3  & Sic1P & y_3 \\
    Clb\cdot Sic1 & x_4 & Sic1\cdot Clb & y_4 \\
    Clb\cdot Sic1P & x_5 & Clb+Sic1 & y_5 \\
    Cdc14 & x_6 & Clb\cdot Sic1 & y_6 \\ 
    Sic1P\cdot Cdc14 & x_7 & Clb & y_7\\
    Clb\cdot Sic1P\cdot Cdc14 & x_8 & Clb+Sic1P & y_8\\
    Sic1 \cdot Clb & x_9 & Clb\cdot Sic1P & y_9 \\
    & & Sic1P+Cdc14 & y_{10} \\
    & & Sic1P\cdot Cdc14 & y_{11} \\
    & & Sic1+Cdc14 & y_{12} \\
    & & Clb\cdot Sic1P + Cdc14 & y_{13} \\
    & & Clb\cdot Sic1P\cdot Cdc14 & y_{14} \\
    & & Clb\cdot Sic1 + Cdc14 & y_{15} \\ \hline
  \end{array}
\end{displaymath}

\subsection{Ordinary differential equations}
\label{sec:bin_com_ODEs}

\begin{subequations}
  \begin{align*}
    \dot x_1 &= k_{1}-k_{2}\, x_{1} + k_{4}\, x_{9}-k_{5}\, x_{1}\,
    x_{3}-k_{6}\, x_{1}\, x_{3} + k_{7}\, x_{4} + k_{15}\, x_{7} \\
    \dot x_2 &= -k_{3}\, x_{2} + k_{9}\, x_{9}-k_{10}\, x_{2}\, x_{3}
    + k_{11}\, x_{5}-k_{13}\, x_{2}\, x_{6} + k_{14}\, x_{7} \\ 
    \dot x_3 &= k_{4}\, x_{9}-k_{5}\, x_{1}\, x_{3}-k_{6}\, x_{1}\,
    x_{3} + k_{7}\, x_{4} + k_{8}\, x_{4} \\
    &\qquad + k_{9}\, x_{9}-k_{10}\,
    x_{2}\, x_{3} + k_{11}\, x_{5} + k_{12}\, x_{5} \\ 
    \dot x_4 &= k_{6}\, x_{1}\, x_{3}-k_{7}\, x_{4}-k_{8}\, x_{4} +
    k_{18}\, x_{8} \\
    \dot x_5 &= k_{10}\, x_{2}\, x_{3}-k_{11}\, x_{5}-k_{12}\,
    x_{5}-k_{16}\, x_{5}\, x_{6} + k_{17}\, x_{8} \\
    \dot x_6 &= -k_{13}\, x_{2}\, x_{6} + k_{14}\, x_{7} + k_{15}\,
    x_{7}-k_{16}\, x_{5}\, x_{6} + k_{17}\, x_{8} + k_{18}\, x_{8} \\
    \dot x_7 &= k_{13}\, x_{2}\, x_{6}-k_{14}\, x_{7}-k_{15}\, x_{7} \\
    \dot x_8 &= k_{16}\, x_{5}\, x_{6}-k_{17}\, x_{8}-k_{18}\, x_{8} \\
    \dot x_9 &= -k_{4}\, x_{9} + k_{5}\, x_{1}\, x_{3}-k_{9}\, x_{9}
  \end{align*}
\end{subequations}

\subsection{Conservation relations}

\begin{align*}
Z_1^Tx= & \ \ x_{6} + x_{7} + x_{8}\ \ = c_1 \\
Z_2^Tx= & x_{3} + x_{4} + x_{5} + x_{8} + x_{9} = c_2
\end{align*}

\subsection{The stoichiometric matrix:}

\begin{center}
  \scalebox{0.5}{
    \begin{minipage}{1.0\linewidth}
      \begin{displaymath}
        S = 
        \left[
          \begin{array}{rrrrrrrrrrrrrrrrrr}
            1 & -1 & 0 & 1 & -1 & -1 & 1 & 0 & 0 & 0 & 0 & 0 & 0 & 0 & 1 & 0 & 0 & 0 \\
            0 & 0 & -1 & 0 & 0 & 0 & 0 & 0 & 1 & -1 & 1 & 0 & -1 & 1 & 0 & 0 & 0 & 0 \\
            0 & 0 & 0 & 1 & -1 & -1 & 1 & 1 & 1 & -1 & 1 & 1 & 0 & 0 & 0 & 0 & 0 & 0 \\
            0 & 0 & 0 & 0 & 0 & 1 & -1 & -1 & 0 & 0 & 0 & 0 & 0 & 0 & 0 & 0 & 0 & 1 \\
            0 & 0 & 0 & 0 & 0 & 0 & 0 & 0 & 0 & 1 & -1 & -1 & 0 & 0 & 0 & -1 & 1 & 0 \\
            0 & 0 & 0 & 0 & 0 & 0 & 0 & 0 & 0 & 0 & 0 & 0 & -1 & 1 & 1 & -1 & 1 & 1 \\
            0 & 0 & 0 & 0 & 0 & 0 & 0 & 0 & 0 & 0 & 0 & 0 & 1 & -1 & -1 & 0 & 0 & 0 \\
            0 & 0 & 0 & 0 & 0 & 0 & 0 & 0 & 0 & 0 & 0 & 0 & 0 & 0 & 0 & 1 & -1 & -1 \\
            0 & 0 & 0 & -1 & 1 & 0 & 0 & 0 & -1 & 0 & 0 & 0 & 0 & 0 & 0 & 0 & 0 & 0
          \end{array}
        \right]
      \end{displaymath}
    \end{minipage}
  }
\end{center}

\subsection{The vector of reaction rates:}
\begin{displaymath}
  \begin{split}
    v(k,x) &= 
    \bigl(
    k_{1},\,
    k_{2} x_{1},\,
    k_{3} x_{2},\,
    k_{4} x_{9},\,
    k_{5} x_{1} x_{3},\,
    k_{6} x_{1} x_{3},\,
    k_{7} x_{4},\,
    k_{8} x_{4},\,
    k_{9} x_{9},\,
    k_{10} x_{2} x_{3},\\
    &\qquad
    k_{11} x_{5},\,
    k_{12} x_{5},\,
    k_{13} x_{2} x_{6},\,
    k_{14} x_{7},\,
    k_{15} x_{7},\,
    k_{16} x_{5} x_{6},\,
    k_{17} x_{8},\,
    k_{18} x_{8}
    \bigr)^T
  \end{split}
\end{displaymath}


\vspace{5mm}
\section{Some Linear Algebra}
\label{sec:imequal}\mbox{}

\subsection{The hypothesis  ${\bf im(S) = im(J(\nu ,x))}$ in (\ref{eq:imequal})}\label{sec:imequal1}
\begin{subequations}
We consider the factorization 
$J=S\, [\YL V]^T\, \diag{(x^{-1})}$ from (\ref{eq:def_Jac_ss})\&(\ref{eq:def_Jac_ssh})
where $V=\diag{(E\nu)}$ and $\diag{(x^{-1})}$ are diagonal matrices with positive entries. 
The equality $im(S) = im(J)$
is thus equivalent to $im(S) = im(S[\YL V]^T)$ since
the invertible factor $\diag{(x^{-1})}$ can be discarded.

Given such $(n\times r)$-matrices $S$ and $B=\YL V$ of $\rank{(S)}=s$
and $\rank{(B)}=rank{(Y)}=:\beta$  respectively, one always has
$\im{(S)}\supset\im{(SB^T)}$ and $s \geq \rank{(SB^T)}$.
We discuss the equality 
\begin{equation}\label{fred1}
  S(\R^r)=\im{(S)}=\im{(SB^T)}=S(\im{B^T})
\end{equation}
which is obviously equivalent to $s=\rank{(SB^T)}=\rank{(BS^T)}$ and to
\begin{equation}\label{fred3a}
  \dim\left(\ker(SB^T)\right) = n-s = \dim\left(\ker(BS^T)\right)
  \, .
\end{equation}
Obviously, (\ref{fred1}) necessitates $\beta \geq s$.
Moreover one has $\im{(S)}\subset\im{(SB^T)}$ if and only if 
$\big[\im{(S)}\big]^{\perp} = \ker{(S^T)}\supset \ker{(B^TS)}=\big[\im{(SB^T)}\big]^{\perp}$ and thus if and only if 
\begin{equation}\label{fred3alt}
BS^T\xi=0\ \Rightarrow \ S^T\xi =0\, .
\end{equation}
The elements $\xi$ of the $(n-s)$-dimensional subspace $\ker{(S^T)}$ satisfy
(\ref{fred3alt}). Therefore, $\rank{(S)}=\rank{(SB^T)}$ is equivalent to
$B\mid_{\im{(S^T)}}$ being an injective map.
This can be reformulated in terms  of matrices $S_c$ and $B_c$, 
whose columns form a basis of $\im{(S^T)}$ and $\im{(B^T)}$
respectively: 
$\rank{(S)}=\rank{(SB^T)}$ is equivalent to 
$B_c^TS_c\in \R^{r
\times s}$ being of full column   rank $s$.
Finally, this fact  leads with the help of matrices $S_c$ and $\YL_c$, whose columns form a basis of 
$\im{(S^T)}$ and $\im{(\YL^T)}$
respectively, to the following characterization: 
\begin{equation}\label{fred8bb}\im{(S)}=\im{(J(\nu ,x))} \ \Leftrightarrow \
\YL_c^T\diag{(E\nu)}\, S_c \in \R^{r\times s} \  \mbox{ has full column  rank }\,  s\, .
\end{equation}
\end{subequations}

\subsection{The reduced system  ${\bf Q(z)\nu=0}$ in
  (\ref{eq:Qz_condi})}\label{sec:imequal2} 

\begin{subequations}
  When discussing $H(z)\, \nu = 0$ with $H(z)$ given by
  (\ref{eq:def_Jac_sshg}), we have introduced matrices $S_0$ and
  $S_c$ such that the columns of $S_0$ and  $S_c$ form a basis of
  $\ker(S)$ and  $\im(S^T)$ respectively. Furthermore we have chosen
  the Moore-Penrose inverse $S_0^\#$ and a  particular matrix $S_\#$
  with $S_\#\, S_0 = \left[ 
    \begin{smallmatrix}
      I_{r-s} \\ {\bf 0}_{s\times (r-s)}
    \end{smallmatrix}
  \right]$, namely  
  $S_\#^{part}=\left[
    \begin{smallmatrix}
      S_0^\#\\ S_c^T
    \end{smallmatrix}
  \right]$.
  Theses choices have led to the equivalence of $H(z)\, \nu = 0$ and
  $Q(z)\nu=0$  with the corresponding $\alpha$ will be given by
  $P(z)\nu$ (cf. the set-up for (\ref{eq:Ssharp0}), (\ref{eq:Teq}) and
  (\ref{eq:Qz_condi})). 
  
  When considering  a different basis representation
  $\tilde{S}_0=S_0R_0$ of  $\ker(S)$ for a regular matrix $R_0\in 
  \R^{(r-s)\times (r-s)}$ and when working with the general form of
  $S_\#$ given by 
  \begin{equation}
    \label{eq:Ssharp1}
    S_\#^{gen}=  \left[\begin{matrix}
        \tilde{S}_0^\#+\Lambda S_c^T\\ R_c^TS_c^T
      \end{matrix}
    \right] =  
    \left[\begin{matrix}
        R_0^{-1} &\Lambda \\0& R_c^T
      \end{matrix}
    \right]\, \left[\begin{matrix} S_0^\#\\ S_c^T\end{matrix}\right]
  \end{equation}
  for regular matrices $R_c\in \R^{s\times s}$ and arbitrary matrices
  $\Lambda\in \R^{(r-s)\times s}$, one arrives at the equivalence of 
  $\, \tilde{S}_0\, \tilde{\alpha} =\diag\left(\YL^T\, z\right)E\, \nu \, $
  (cf.~(\ref{eq:Ssharp0})) to
  \begin{equation}
    \label{eq:Ssharp2}
    \left[\begin{matrix}
        R_0^{-1} &\Lambda \\0& R_c^T
      \end{matrix}
    \right]\, \left[\begin{matrix} P(z)\, \nu\\ Q(z)\, \nu\end{matrix}\right]\ = \
    \left[\begin{matrix}\tilde{\alpha}\\0
      \end{matrix}
    \right]
  \end{equation}
  for some matrix $\Lambda$ of suitable dimensions.
    Hence, for general $S_\#^{gen}$ one obtains (in analogy to
    (\ref{eq:Qz_condi})) the condition 
    \begin{equation}
      \label{eq:gen_Qz_nu}
      R_c^T\, Q(z)\, \nu = 0.
    \end{equation}
    As $R_c$ is regular, one has that any pair ($z$, $\nu$)$\in
    \R^n\times {\cal V}$ satisfying (\ref{eq:gen_Qz_nu}) also satisfies
    (\ref{eq:Qz_condi}) and vice versa. Hence we conclude that
    (\ref{eq:Qz_condi}) is independent from the chosen bases for
    $\ker(S)$ and $\im(S^T)$. 
The corresponding $\tilde{\alpha}$ 
depends on the choice of $R_0$ and $\Lambda$
as (\ref{eq:Ssharp2}) shows. 
\end{subequations}

\subsection{Ad Remark~\ref{openeq:Teq}}
\label{sec:imequal3} 

By Appendix~\ref{sec:imequal1} one has $\im{(S)}=\im{(J)}$ if and only
if one of the $(s\times s)$-minors of $\YL_c^T\diag{(E\nu)}S_c$ is
nonzero for the chosen $\nu$ from the kernel of $Q(z)$. We like to add
that these minors are polynomials in the components of $E\nu$ of order
not greater than $s$. By Remark~\ref{openeq:Teq} following
Theorem~\ref{theo:nasc}, the $\nu$'s  might be varied locally. Such a
variation might be employed to establish (\ref{fred8bb}).



\end{document}